\documentclass[sigconf,balance=false]{acmart}

\usepackage{popets}
	\setcopyright{popets}
	\copyrightyear{YYYY}
	\acmYear{YYYY}
	\acmVolume{YYYY}
	\acmNumber{X}
	\acmDOI{XXXXXXX.XXXXXXX}
	\acmISBN{}
	\acmConference{Proceedings on Privacy Enhancing Technologies}
\usepackage{thmtools,thm-restate}
\usepackage{subcaption} 
\usepackage{tablefootnote}
\usepackage{algorithm}
\usepackage[noend]{algpseudocode}

\usepackage{siunitx}
\usepackage{tikz}
\newcommand*{\priority}[1]{\begin{tikzpicture}[scale=0.09]%
    \draw (0,0) circle (1);
    \fill[fill opacity=0.99,fill=black] (0,0) -- (90:1) arc (90:90-#1*3.6:1) -- cycle;
    \end{tikzpicture}}

\newtheorem{definition}{Definition}[section]

\newtheorem{example}{Example}

\newtheorem{remark}{Remark}
\newtheorem{observation}{Observation}

\usepackage{dcolumn}
\newcolumntype{.}{D{.}{.}{-1}}

\newcommand{\tablespace}{\noalign{\vskip 2mm}}

\usepackage[nonumberlist,acronyms,nogroupskip]{glossaries}
\glsdisablehyper

\newacronym{HE}{HE}{Homomorphic Encryption}
\newacronym{MPC}{MPC}{Multi Party Computations}
\newacronym{NN}{NN}{Neural Network}
\newacronym{SIMD}{SIMD}{Single Instruction Multiple Data}
\newacronym{RGB}{RGB}{Red, Green, Blue}
\newacronym{FC}{FC}{Fully Connected}
\newacronym{RMSE}{RMSE}{Root Mean Square Error}

\newcommand{\floor}[1]{\left\lfloor {#1} \right\rfloor}
\newcommand{\ceil}[1]{\left\lceil {#1} \right\rceil}

\newcommand{\BigO}{\mathcal{O}}

\newcommand{\strideh}{\delta_h}
\newcommand{\stridew}{\delta_w}

\renewcommand{\paragraph}[1]{{\medskip\noindent {\bf{#1}}~}}

\begin{document}

\title[HeLayers: A Tile Tensors Framework for Large Neural Networks on Encrypted Data]{\huge HeLayers: A Tile Tensors Framework for Large Neural Networks on Encrypted Data}

\settopmatter{authorsperrow=5}
\author{Ehud Aharoni}
\orcid{0000-0002-3647-1440}
\affiliation{IBM Research \country{Israel}}

\author{Allon Adir}
\orcid{0000-0001-8128-6706}
\affiliation{IBM Research \country{Israel}}

\author{Moran Baruch}
\orcid{0000-0003-0615-6164}
\affiliation{IBM Research \country{Israel}}
\affiliation{Bar Ilan University \country{Israel}}

\author{Nir Drucker}
\orcid{0000-0002-7273-4797}
\affiliation{IBM Research \country{Israel}}

\author{Gilad Ezov}
\orcid{0000-0003-4579-8127}
\affiliation{IBM Research \country{Israel}}

\author{Ariel Farkash}
\orcid{0000-0003-1487-9223}
\affiliation{IBM Research \country{Israel}}

\author{Lev Greenberg}
\orcid{0000-0002-1981-9775}
\affiliation{IBM Research \country{Israel}}

\author{Ramy Masalha}
\orcid{0000-0002-6808-5675}
\affiliation{IBM Research \country{Israel}}

\author{Guy Moshkowich}
\orcid{0000-0003-1856-8430}
\affiliation{IBM Research \country{Israel}}

\author{Dov Murik}
\orcid{0000-0001-6366-8882}
\affiliation{IBM Research \country{Israel}}

\author{Hayim Shaul}
\orcid{0000-0001-8432-0623}
\affiliation{IBM Research \country{Israel}}

\author{Omri Soceanu}
\orcid{0000-0002-7570-4366}
\affiliation{IBM Research \country{Israel}}

\renewcommand{\shortauthors}{}

\begin{abstract}
Privacy-preserving solutions enable companies to offload confidential data to third-party services while fulfilling their government regulations. 
To accomplish this, they leverage various cryptographic techniques such as Homomorphic Encryption (HE), which allows performing computation on encrypted data. Most HE schemes work in a SIMD fashion, and the data packing method can dramatically affect the running time and memory costs. Finding a packing method that leads to an optimal performant implementation is a hard task. 
We present a simple and intuitive framework that abstracts the packing decision for the user. We explain its underlying data structures and optimizer, and propose a novel algorithm for performing 2D convolution operations.
We used this framework to implement an inference operation over an encrypted HE-friendly AlexNet neural network with large inputs, which runs in around five minutes, 
several orders of magnitude faster than other state-of-the-art non-interactive HE solutions. 
\end{abstract}

\maketitle
    
\section{Introduction}\label{sec:intro}

\gls{HE} schemes allow computations to be performed over encrypted data while providing data confidentiality for the input. Specifically, they allow the evaluation of  functions on encrypted input, which is useful when outsourcing sensitive data to a third-party cloud environment. For example, a hospital that provides an X-ray classification service (e.g., COVID-19 versus pneumonia) can encrypt the images using \gls{HE}, express the classification algorithm as a function, and ask a cloud service to evaluate it over the encrypted data without decrypting it. In this way, the hospital can use the cloud service while possibly complying with regulations such as HIPAA~\cite{HIPAA} and GDPR~\cite{GDPR}.

The proliferation of \gls{HE} solutions in the last decade \cite{gartner} shows that customers are eager to use them and that companies and organizations strive to provide secure and efficient solutions such as HEBench \cite{hebench}. Nevertheless, it turns out that running large \glspl{NN} using \gls{HE} only is still considered an expensive task. For example, the best implementations \cite{REDsec} of AlexNet \cite{AlexNet2012} with large inputs of $224 \times 224 \times 3$ before this paper, was measured as taking more than $3$ hours using only $80$ security bits. 
This barrier forces users to search for other secure alternatives instead of enjoying the advantage of solutions that rely only on \gls{HE}. Our proposed framework aims to narrow down this barrier, allowing the users to better utilize cloud capabilities while operating on their confidential data.

Some \gls{HE} schemes, such as CKKS~\cite{CKKS2017}, operate on ciphertexts in a homomorphic \gls{SIMD} fashion. This means that a single ciphertext encrypts a fixed size vector, and the homomorphic operations on the ciphertext are performed slot-wise on the elements of the plaintext vector.
To utilize the \gls{SIMD} feature, we need to pack and encrypt more than one input element in every ciphertext. The  packing method can dramatically affect the {\em latency} (i.e., time to perform computation), {\em throughput} (i.e., number of computations performed in a unit of time), communication costs, and memory requirements.
The designer of an HE-based system is usually tasked with choosing the ciphertext-size. The size is bounded from above by the FHE-library implementation and from below by security parameters. Once selected, the ciphertext-size stays fixed for the entire process (with different packing options).
To demonstrate the effect of different packing choices we use CryptoNets~\cite{CryptoNets2016}. We summarize the results in Table~\ref{tab_inference1}, and observe that using two na\"ive packing solutions achieved latencies of 0.86 sec. and 11.1 sec., with memory usage of 1.58 GB and 14 GB, respectively. In comparison, a different non-trivial packing method achieved better latency of 0.56 sec. and memory usage of 0.73 GB. Section \ref{section:nn} provides more details about these three packing methods.

We also demonstrate why packing is hard by presenting the problems involved in encrypting and packing a matrix. A na\"ive packing option may keep every row or column in a different ciphertext. When the evaluated algorithm performs manipulations over rows, one packing option is clearly  better than the other. However, when the matrix dimensions are small, the encrypted ciphertext may include many unused ciphertext slots. With the cost of performing extra rotations, it is possible to pack more than one row in a single ciphertext.

Designing a good packing method is not straightforward (e.g., \cite{low_depth_circuit, GAZELLE2018, GALA}) and the most efficient packing method may not be the trivial one (see above). Moreover, different optimization goals may lead to different packing, e.g., as shown in Table~\ref{tab_inference2}. As the size of the \gls{HE} code increases, it becomes harder to find the optimal packing. For example, finding the best packing for a large \gls{NN} inference algorithm is hard since the input is typically a four or five dimensional tensor, and the computation involves a long sequence of operations such as matrix multiplication and convolution.

\paragraph{Non-interactive HE.} Two approaches for \gls{HE} computations are client-aided and non-client-aided. In the client-aided approach, during the computation the server asks the data owner or the end-user for assistance. I.e., the user is asked to decrypt the intermediate ciphertext results, perform some minor computation tasks, and re-encrypt the data using \gls{HE}. This approach was implemented in GAZELLE \cite{GAZELLE2018} and NGraph \cite{HET} using \gls{MPC}. It 
has the drawback that the client must stay online during the computation. Moreover, this setting poses some security concerns~\cite{AkaviaVald21} or can even make it easier to perform model-extraction attacks \cite{muse}. To avoid these limitations, we focus on the non-client-aided approach where the computation is done entirely under encryption, without interaction.

Using non-client-aided designs require using \gls{HE}-friendly \gls{NN} architectures; these replace nonlinear layers such as ReLU and MaxPooling with other functions. Today, these conversions have become common practice (see survey \cite{sok}), and they present a time versus accuracy tradeoff that is mostly analyzed in AI-related works (e.g., \cite{baruch2021fighting}).  Our framework can work with any polynomial activation function. Hence, our current and future users can decide how to balance time and accuracy by choosing the HE architecture that best suits their needs. We leave the accuracy discussion outside the scope of this paper. Here, we emphasize that our security-oriented framework leverages the \gls{SIMD} property of \gls{HE}-schemes, which can independently benefit from any AI-domain improvement in model accuracy. 

Using HE-friendly models may require retraining models before using them. Nevertheless, we argue that the proliferation of the HE-domain, together with future AI improvements, will bring about more solutions that prefer training HE-friendly models directly, from day one. With our focus on non-client-aided designs, we provide customers with better security guarantees and improved client usability. The potential small cost in model accuracy is expected to  get smaller over time. 

Some recent \gls{HE} compilers \cite{chet_compiler, HET} simplify the way users can implement \gls{NN} solutions on encrypted data by allowing them to focus on the network and leaving the packing optimizations to the compilers. This is also the purpose of our tile tensor framework. It enables us to evaluate an \gls{HE}-friendly version  \cite{baruch2021fighting} of AlexNet \cite{AlexNet2012} in only five minutes. To the best of our knowledge, this is the largest network to be implemented with a feasible running time, 128-bit security, in a non client-aided mode, and without bootstrapping (see \cite{Halevi2017}). In comparison, NGraph \cite{HET} reported their measurements for CryptoNets \cite{CryptoNets2016} or for MobileNetV2 \cite{mobilenetv2} when using client-aided design, and CHET \cite{chet_compiler} reported the results for SqueezeNet \cite{squeezenet}, which has $50\times$ fewer parameters than AlexNet. Another experiment using NGraph and CHET was reported in \cite{sok} using Lenet-5 \cite{lenet-5}, which is also a small network compared to AlexNet. See a complete comparison at Section \ref{section:comparison}.

\subsection{Our Contribution}
Our contributions can be summarized as follows:
\begin{itemize}
    \item {\em A tile tensor based framework}. We introduce a new packing-oblivious programming-framework that allows users to concentrate on the \gls{NN} design instead of the packing decisions. This framework is simple and intuitive, and is available for non-commercial use in \cite{helayersdocker}.

    \item {\em Tile tensors}. We introduce tile tensors, a new generalized packing method that intuitively covers many of the previous packing optimizations. A tile tensor offers the user the API of an ordinary tensor, and is implemented using elegant algorithms.

    \item {\em Packing optimizer}. We describe a packing optimizer that considers many different packing options. The optimizer estimates the time and memory needed to run a given function for every option, and reports the one that performs best per a given objective, whether latency, throughput, or memory.

    \item {\em A new method to compute convolution}. We provide a new packing method and a new implementation of the 2D convolutional layer, which is a popular building block in \glspl{NN}. Our new packing and implementation are more efficient for large inputs than previous work. In addition, with this packing, we are able to efficiently compute a long sequence of convolution-layers.

    \item {\em Efficient \gls{HE}-friendly version of AlexNet inference under encryption}. We implemented an \gls{HE}-friendly version of AlexNet. 
    To the best of our knowledge, this is the fastest non-client-aided evaluation of this network.
    
    \item {\em Packing notation}. We present a language for describing packing details. It covers several known packing schemes, as well as new ones, and allows easy and intuitive \gls{HE} circuit design.
\end{itemize}

The rest of the paper is organized as follows. 
Section~\ref{section:background} describes the notation used in the paper, and some background terminology. 
Section~\ref{section:ttfw} provides an overview of the tile tensor framework. We describe the tile tensors data structure in Section~\ref{section:tt_intro} and the packing optimizer in Section~\ref{section:optimizer}. Section~\ref{section:convolution} describes our novel convolution algorithm, and Section~\ref{section:nn} reports the results of our experiments when using CryptoNets and AlexNet. In Section~\ref{section:comparison}, we provide an extended comparison of our methods to the state-of-the-art methods. Section~\ref{section:conclusions} concludes the paper. 

\section{Background}
\label{section:background}
\subsection{Notation}
\label{subsection:notation}

We use the term {\em tensor} as synonymous with multi-dimensional array, as  this is common in the AI domain. We denote the shape of a $k$-dimensional tensor by $[n_1,n_2,\ldots,n_k]$, where $0 < n_i$ is the size of the $i$'th dimension. For example, the shape of the $5\times 6$ matrix $M$ is $[5,6]$. We sometimes refer to a tensor $M$ by its name and shape $M[5,6]$ or just by its name $M$ when the context is clear. For a tensor $R[n_1,\ldots,n_k]$, we use $R(j_1,j_2,\ldots,j_k)$ to refer to a specific element, where $0 \leq j_i < n_i$. We use uppercase letters for tensors.

We write matrix multiplication without a multiplication symbol, e.g., $M_1 M_2$ stands for the product of $M_1$ and $M_2$. We denote the transpose operation of a matrix $M$ by $M^T$ and we use tags (e.g., $M',M''$) to denote different objects. The operations $M_1+M_2$ and $M_1*M_2$ refer to element-wise addition and multiplication, respectively.

\subsection{Tensor Basic Operations}
\label{subsection:tensor_basics}
\subsubsection{Broadcasting and Summation}
\label{subsection:broadcasting}

Here we define some commonly used tensor terms and functions.

\begin{definition}[Compatible shapes]
The tensors $A[n_1,\ldots,n_k]$ and $B[m_1,\ldots,m_k]$ have {\em compatible shapes} if $m_i = n_i$ or either $n_i = 1$ or $m_i = 1$, for $i \le k$. Their {\em mutual expanded shape} is $[\max\{n_i, m_i\}]_{i \le k}$.
\end{definition}

\begin{remark}
When a tensor $A$ has more dimensions than a tensor $B$, we can match their dimensions by expanding $B$ with dimensions of size $1$. This results in equivalent tensors up to transposition. For example, both tensors $V[b]$ and $V[b,1]$ represent column vectors, while $V[1,b]=V^T$ represents a row vector.
\end{remark}

The broadcasting operation takes two tensors with compatible but different shapes and expands every one of them to their mutual expanded shape. 

\begin{definition}[Broadcasting]
For a tensor $A[n_1,\ldots,n_k]$ and a tensor shape $s=[m_1,\ldots,m_k]$ with $n_i \in \{1,m_i\}$ for each $i=1,\ldots,k$, the operation $C = broadcast(A, s$) replicates the content of $A$ along the $r$ dimension $m_r$ times for every $r=1,\ldots,k$ and $n_r = 1 < m_r$. The output tensor $C$ is of shape $s$.
\end{definition}

\begin{example} 
The tensors $A[3,4,1]$ and $B[1,4,5]$ have compatible shapes. Their mutual expanded shape is $s=[3,4,5]$ and\\$broadcast(A, s)$ has the same shape $s$ as $broadcast(B, s)$.
\end{example} 

We perform element-wise operations such as addition ($A+B$) and multiplication ($A*B$) on two tensors with compatible shapes $A,B$ by first using broadcasting to expand them to their mutual expanded shape and then performing the relevant element-wise operation.

\begin{definition}[Summation]
For a tensor $A[n_1,\ldots,n_k]$, the operation $B=sum(A,t)$ sums the elements of $A$ along the $t$-th dimension and the resulting tensor $B$ has shape $[n_1,\ldots,n_{t-1},1,\ldots,n_k]$ and
$$B(j_1,\ldots,j_{t-1},0,\ldots,j_k) = \sum_{l=0}^{n_t-1}A(j_1,\ldots,j_{t-1},l,\ldots,j_k),
$$
For all $j_i < n_i$ for $i\in \{1,2,\ldots,k\} \setminus \{t\}$, 

\end{definition}

Using broadcasting and summation we can define common algebraic operators.
For two matrices $M_1[a,b]$, $M_2[b,c]$ and the column vector $V[b,1]$, we compute matrix-vector multiplication using $M_1 V = sum(M_1 * V^T, 2)$, where $M_1$ and $V^T$ have compatible shapes with the mutual expanded shape of $[a,b]$. We compute matrix-matrix multiplication using $M_1 M_2 = sum(M_1' * M_2',2)$, where $M_1'= M_1[a,b,1]$ and $M_2' = M_2[1,b,c]$.

\subsubsection{Convolution}
\label{subsection:convolution}

2D convolution is a popular building block in \glspl{NN}. Its input is often an images tensor $I[w_I, h_I, c, b]$ and a filters tensor $F[w_F, h_F, c, f]$ with the following shape parameters: width $w_I, w_F$, height $h_I, h_F$, and the number of image channels $c$ (e.g., 3 for an RGB image). 
In addition, we compute the convolution for a batch of $b$ images and for $f$ filters. Informally, the convolution operator moves each filter in $F$ as a sliding window over elements of $I$ starting at position $(0,0)$ and using strides of $\stridew$ and $\strideh$. When the filter fits entirely in the input, the dot product is computed between the elements of the filter and the corresponding elements of $I$.

\begin{definition}[Convolution]
Let $I[w_I,h_I,c,b]$ and\\$F[w_F,h_F,c,f]$ be two input tensors for the convolution operator representing images and filters, respectively. The results of the operation $O = conv2d(I, F)$ is the tensor $O[w_O,h_O,f,b]$, where $w_O=\ceil{\frac{w_I - w_F + 1}{\stridew}}$, $h_O = \ceil{\frac{h_I - h_F + 1}{\strideh}}$, $\strideh$ and $\stridew$ are the strides and
\begin{align}
\nonumber O( &i,j,m,n)= \\
& \sum_{k=0}^{w_F-1} \sum_{l=0}^{h_F-1} \sum_{p=0}^{c-1} I(i\cdot\stridew+k,j\cdot\strideh+l,p,n) F(k,l,p,m).
\label{equ:conv_full}
\end{align}
\end{definition}

In the degenerated case where $\stridew=\strideh=b=f=c=1$, Equation \eqref{equ:conv_full} can be simplified to
\begin{equation}
O(i,j)=\sum_{k=0}^{w_F-1} \sum_{l=0}^{h_F - 1} I(i+k,j+l) F(k,l).
\label{equ:conv_one}
\end{equation}

\subsection{Homomorphic Encryption}
\label{subsection:background_he}

An \gls{HE} scheme is an encryption scheme that allows us to evaluate any circuit, and any function, on encrypted data. A survey is available in \cite{Halevi2017}. Modern \gls{HE} instantiations such as CKKS \cite{CKKS2017} rely on the hardness of the Ring-LWE problem, support SIMD operations, and operate over rings of polynomials. They include the following methods $Gen, Enc, Dec, Add, Mul$ and $Rot$ which we now briefly describe.
The function $Gen$ generates a secret key public key pair. The function $Enc$ gets a message that is a vector $M[s]$ and returns a ciphertext. Here, $s$ denotes the number of plaintext vector elements {\em (slot count). It} is determined during the key generation.
The function $Dec$ gets a ciphertext and returns an $s$-dimensional vector. For correctness we require $M = Dec(Enc(M))$. The functions $Add$, $Mul$ and $Rot$ are then defined as
\begin{align*}
Dec(Add(Enc(M),Enc(M'))) &=M + M'\\
Dec(Mul(Enc(M),Enc(M'))) &=M * M'\\
Dec(Rot(Enc(M),n))(i) &=M((i+n) \mod s)\\
\end{align*}
An {\em approximation} scheme, such as CKKS \cite{CKKS2017}, is correct up to some small error term, i.e., $|M(i) - Dec(Enc(m))(i)| \le \epsilon$, for some $\epsilon > 0$ that is determined by the key.

\subsection{Reducing Convolution to Matrix-matrix Multiplication}
\label{subsec:background_im2col}
In \gls{HE} settings it is sometimes useful to convert a convolution operation to a matrix-matrix multiplication by pre-processing the input before encrypting it.
One such method is \emph{image-to-column}~\cite{im2col}, which works as follows for the case $c=b=1$. Given an image $I[w_I,h_I]$ and $f$ filters $F[w_F,h_F,f]$, the operator $I',F'=im2col(I,F)$ computes a matrix $I'[w_O h_O, w_F h_F]$, where each row holds the content of a valid window location in $I$ flattened to a row-vector, and $F'[w_F h_F, f]$ contains every filter of $F$ flattened to a column-vector. Here, the tensor $O'[w_O h_O,f]=I' F'$ is a flattened version of the convolution result $O[w_O,h_O,f]=conv2d(I,F)$.

We propose a variant $I'',F''=im2col'(I,F)$ that computes $I''[w_O h_O f, w_F h_F]$ by consecutively replicating $f$ times every row of $I'$, and $F''[w_O h_O f, w_F h_F]$ by concatenating $w_O h_O$ times the matrix $F'^T$. The tensor $O''[w_O h_O f,1]=sum(I''*F'',2)$ contains the convolution result $O[w_O,h_O,f]$. The advantage of this variant is that the output is fully flattened to a column vector, which is useful in situations where flattening is costly (e.g., in \gls{HE}, where encrypted element permutations are required).
The drawback of the $im2col$ method is that it is impossible to perform two consecutive convolution operators without costly pre-processing in between. Moreover, this pre-processing 
increases the multiplicative-depth, deeming it impractical for networks such as AlexNet, where the multiplicative-depth was already near the underlying \gls{HE}-library's limit. In comparison, our convolution methods do not require pre-processing between two consecutive calls.

\subsection{Threat Model}
Our threat model involves three entities: An AI model owner, a cloud server that performs model inference on HE encrypted data using the pre-computed AI model, and users who send confidential data to the cloud for model inference. Among other, we take into account the following three models: 
a) the users and the model owner belong to the same organization, where both have access to the private key and are allowed to see the encrypted data; 
b) the user is the private-key owner. The model owner can use the public key to encrypt its model before uploading it to the cloud. This scenario involves a non-collusion assumption between the user and the cloud; 
c) the model-owner is the private-key holder. The user uses the public key to encrypt its samples before uploading them to the cloud. The model owner can decrypt and distribute the inference or post-processed results to the users. This scenario involves a non-collusion assumption between the model-owner and the cloud. In all cases, the cloud should learn nothing about the underlying encrypted data. In Scenarios b and c the users should not learn the model excluding privacy attacks, where the users try to extract the model training data through the inference results.
We assume that communications between all entities are encrypted using a secure network protocol such as TLS 1.3, i.e., a protocol that provides confidentiality, integrity, and allows the model owner and the users to authenticate the cloud server. The model owner can send the model to the server either as plaintext or encrypted. When the model is encrypted, the server should learn nothing about the model but its structure. In both cases, we assume that the cloud is semi-honest (or honest-but-curious), i.e., that it evaluates the functions provided by the data owner and users without any deviation.
We stress that our packing methods modify the data arrangement before encrypting it and thus do not affect the semantic security properties of the underlying HE scheme.
Finally, in our experiments we target an \gls{HE} solution with $128$-bit level  security.
\section{Our Tile Tensor Framework}
\label{section:ttfw}

\gls{HE} libraries such as HElib \cite{HaleviShoup2014helib} and SEAL \cite{sealcrypto} provide simple APIs for their users (e.g., encrypt, decrypt, add, multiply, and rotate). Still, writing an efficient program that involves more than a few operations is not always straightforward.

\begin{figure}[ht!]
    \centering
    \includegraphics[width=0.7\linewidth]{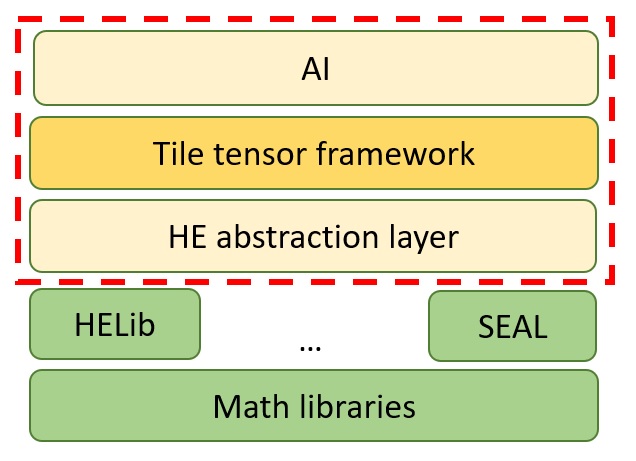}
   \caption{A simplified schematic of the layers in our library. }
  \label{fig:lib}
\end{figure}

Providing users with the ability to develop complex and scalable \gls{HE}-based programs is the motivation for developing higher-level solutions such as our library, NGraph \cite{HET}, and CHET \cite{chet_compiler}. These solutions rely on the low-level \gls{HE} libraries while offering additional dedicated optimizations, such as accelerating \glspl{NN} inference on encrypted data. 

Figure \ref{fig:lib} provides a simplified schematic view of the layers we use in our library. 
The first two layers include the low-level \gls{HE} libraries and their underlying software and hardware math accelerators. 
Our library \cite{helayersdocker} involves the three upper layers.
At the bottom of these layers is the {\em HE abstraction layer}, which makes our library agnostic to the underlying \gls{HE} library.
The next layer is the {\em tile tensor framework layer}. It contains the tile tensor data structure (Section \ref{section:tt_intro}) that simplifies computation involving tensors, and the packing optimizer (Section \ref{section:optimizer}) that searches for the most efficient tile tensor packing configuration for a given computation. The AI layer is built on top of these. It implements AI related functionality, such as reading \glspl{NN} from standard file formats, encrypting their weights, and computing inference. 

In this paper we focus on the tile tensor framework layer, and how it contributes to the optimization of \gls{NN} inference computations. The optimizations this layer offers focus on packing and related algorithms, and can thus be combined with other types of optimizations in other layers. Note that our optimizer (Section \ref{section:optimizer}) is simulation based, and therefore can take into account optimizations in any layer below this layer. 
We refer the reader to \cite{hebase} for APIs and code of the \gls{HE} abstraction layer.

\section{Tile Tensors}
\label{section:tt_intro}

The tile tensor framework layer uses our new data structure, which named tile tensor.
A tile tensor is a data structure that packs tensors in fixed size chunks, as required for \gls{HE}, and allows them to be manipulated similar to regular tensors. It has an accompanying notation for describing the packing details.

We briefly and informally describe both data structure and notation. The notation is extensively used in the rest of the paper, and it is summarized for quick reference in Table~\ref{tab:notation}.
For completeness, we provide formal definitions in Appendix~\ref{appendix:tt_def}. 

\subsection{Tiling Basics}
\label{subsec:tiling}
We start by describing a simple tiling process in which we take a tensor $A[n_1,n_2,\ldots,n_k]$, and break it up into equal-size blocks that we call {\em tiles}, each having the shape $[t_1,t_2,\ldots,t_k]$. In our implementation, a tile is an \gls{HE} ciphertext.

\begin{table}[ht!]
\begin{center}
\caption{Tile tensor shape notation.}
\label{tab:notation}
\renewcommand{\arraystretch}{1.2}
\begin{tabular}{cl|cl}
\hline
$\frac{n_i}{t_i}$ & Basic tiling &
$n_i$ & Basic tiling, $t_i=1$\\
\hline
$\frac{*}{t_i}$ & Replication, $n_i=1$ &
$\frac{n_i?}{t_i}$ & Unknown values\\
\hline
$\frac{n_i \sim}{t_i}$ & Interleaved tiling & $\frac{n_i \sim e_i}{t_i}$ & Interleaved, given $e_i$\\
\hline
\end{tabular}
\end{center}
\end{table}

We construct a tensor $E[e_1,e_2,\ldots,e_k]$, which we call the {\em external tensor}, such that each element of $E$ is a tile, and $e_i=\ceil{\frac{n_i}{t_i}}$.  Thus, $T=E(a_1,a_2,\ldots,a_k)$ for $0\leq a_i < e_i$, is a specific tile in $E$, and $T(b_1,b_2,\ldots,b_k)$ for $0\leq b_i < t_i$ is a specific slot inside this tile.
An element of the original tensor $A(c_1,c_2,\ldots,c_k)$ will be mapped to tile indices $a_i=\floor{\frac{c_i}{t_i}}$, and indices inside the tile $b_i=c_i\mod t_i$. All other slots in $E$ that were not mapped to any element of $A$ will be set to $0$.
Figure~\ref{fig:MM56} demonstrates three examples for tiling a matrix $M[5,6]$.
In Figure~\ref{fig:M5_2__6_4}, for example, the shape of the external tensor is $[3,2]$, and the tile shape is $[2,4]$.

\subsection{The Tile Tensor Data Structure}

A tile tensor is a data structure containing an external tensor as described above, and \textit{public} meta data called {\em tile tensor shape}. The tile tensor shape defines the shape of the tiles, the shape of the original tensor we started with, and some additional  details about the organization of data inside the external tensor, which we describe later.

\begin{figure}[ht]
\centering
\begin{subfigure}{0.48\linewidth}
    \centering
    \includegraphics[scale=0.24]{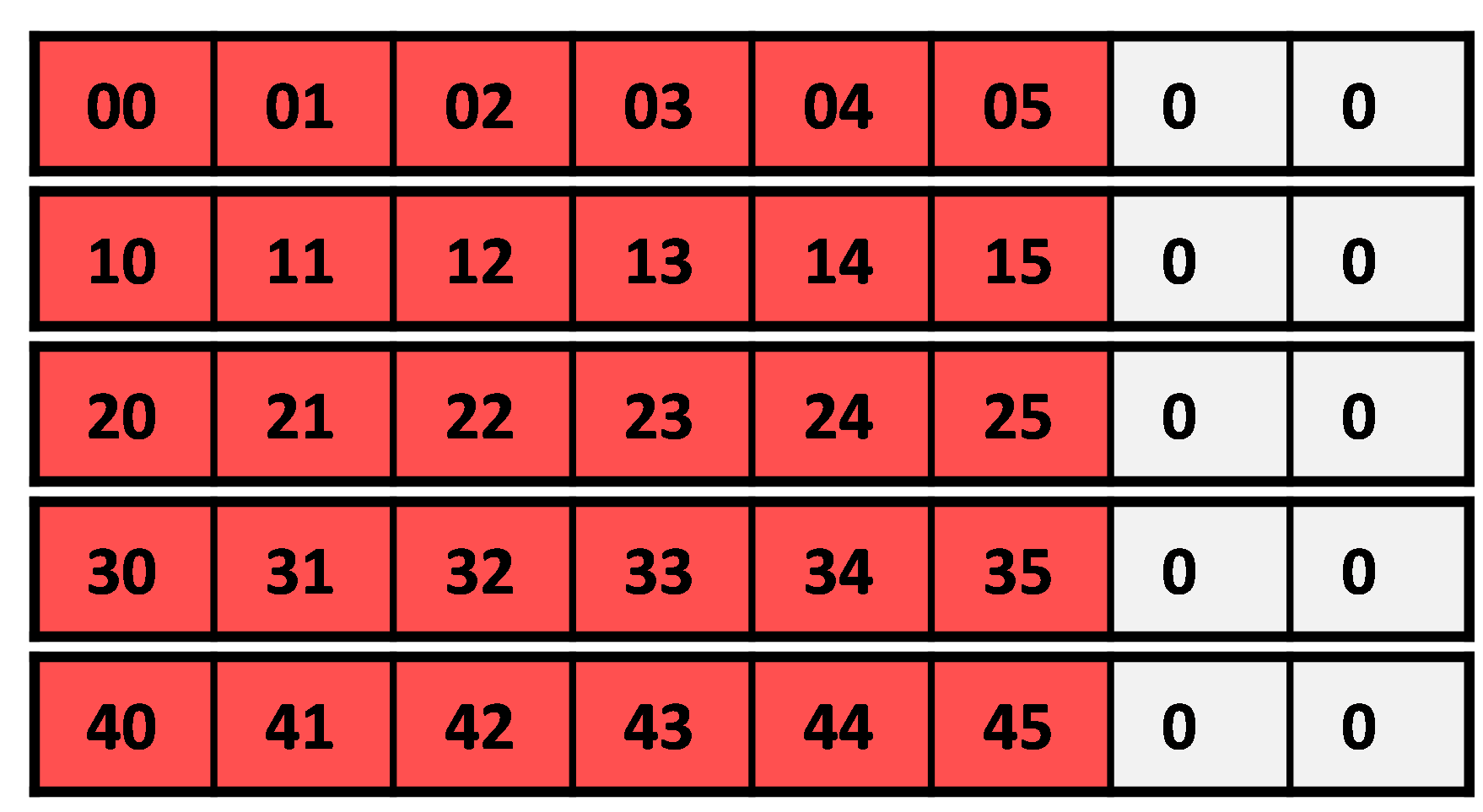}
    \caption{{$T_M[5,\frac{6}{8}]$}}
    \label{fig:M5__6_8}
  \end{subfigure}
    \hfill
  \begin{subfigure}{0.49\linewidth}
    \centering
     \includegraphics[scale=0.24]{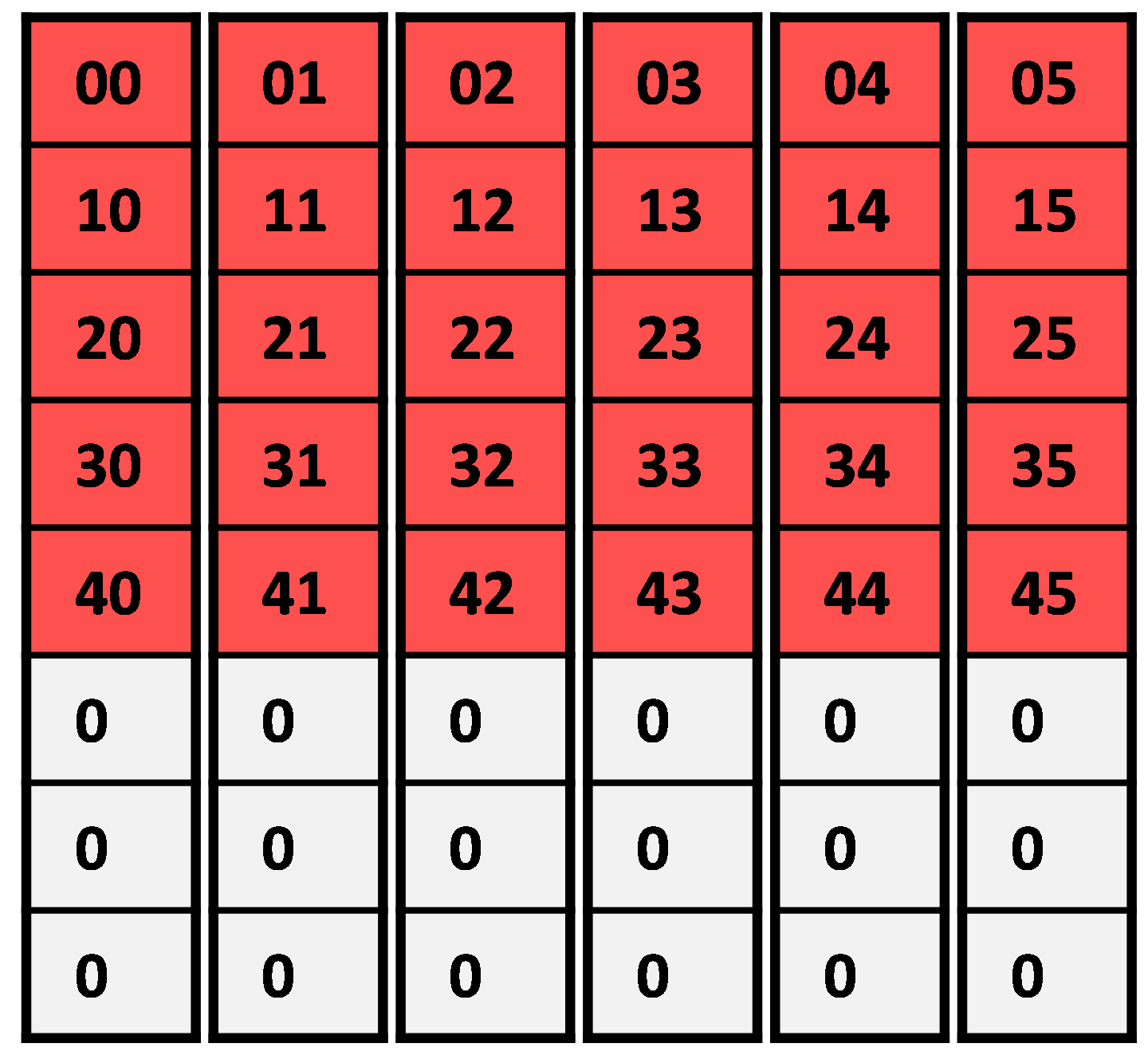}
    \caption{{$T_M'[\frac{5}{8}, 6]$}}
    \label{fig:M5_8__6}
  \end{subfigure}
  \par\bigskip
  \par\bigskip
  \begin{subfigure}{0.5\linewidth}
    \centering
    \vspace{-12pt}
    \includegraphics[scale=0.24]{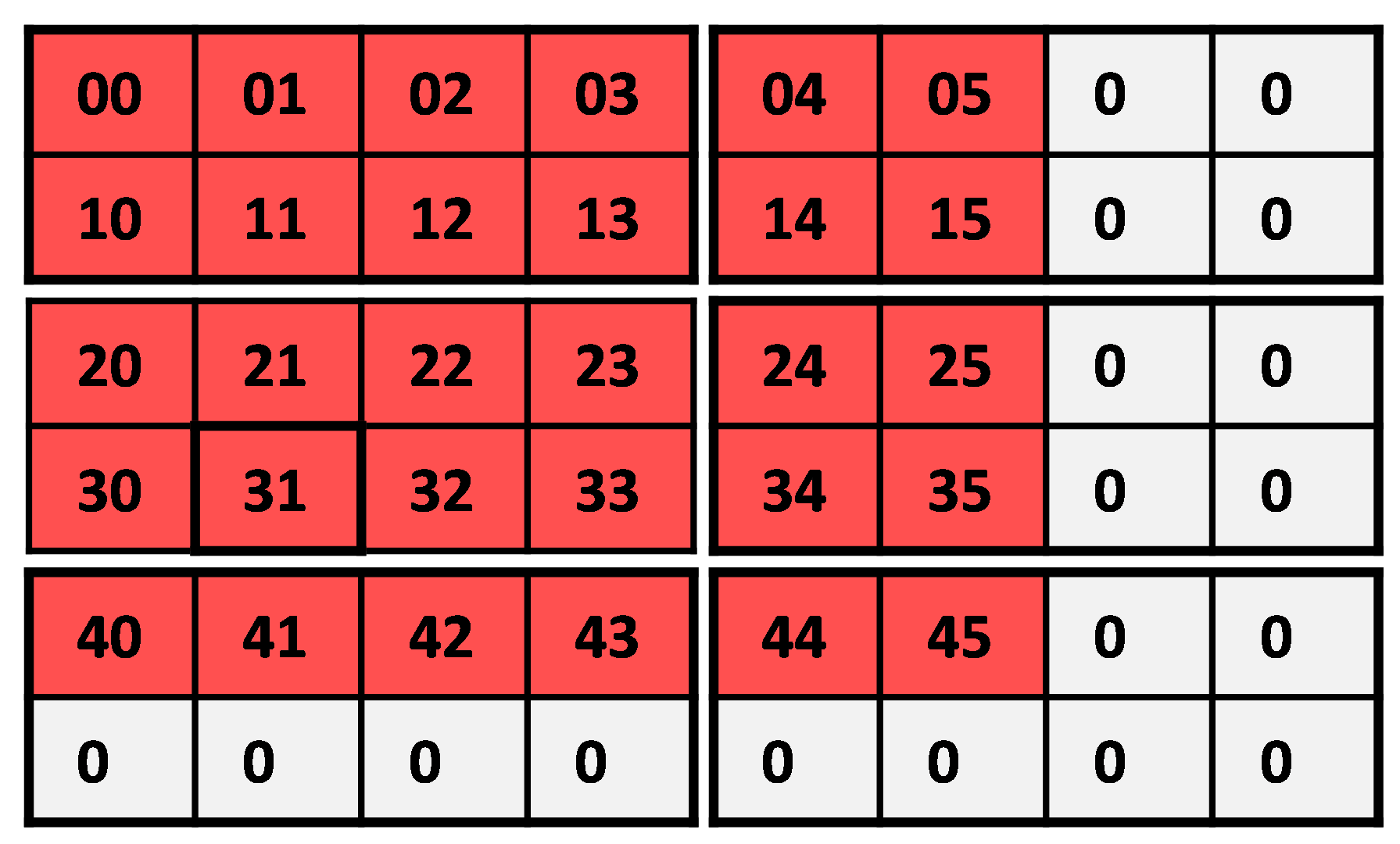}
    \caption{{$T_M''[\frac{5}{2},\frac{6}{4}]$}}
    \label{fig:M5_2__6_4}
  \end{subfigure}
  \caption{Packing an $M[5,6]$ tensor into the tile tensors $T_M$ (Panel a), $T'_M$ (Panel b), $T''_M$ (Panel c) with $8$-slot tiles of shape $[1,8]$, $[8,1]$, and $[2,4]$, respectively. For these, the external tensor shape is $[5,1]$, $[1,6]$, and $[3,2]$, respectively. The value of $M[5,6]$ in the $i$th row and $j$th column is the value $ij$.}
  \label{fig:MM56}
\end{figure}

We use a special notation to denote tile tensor shapes. For example,  $[\frac{n_1}{t_1},\frac{n_2}{t_2},\ldots,\frac{n_k}{t_k}]$ is a tile tensor shape specifying that we started with a tensor of shape $[n_1,\ldots,n_k]$ and tiled it using tiles of shape $[t_1,\ldots,t_k]$.
In this notation, if $t_i=1$, then it can be omitted. For example, $[\frac{5}{1},\frac{6}{8}]$ can be written $[5,\frac{6}{8}]$. 
Figure~\ref{fig:MM56} shows three examples along with their shapes.

A tile tensor can be created using a \emph{pack} operation that receives a tensor $A[n_1,\ldots,n_k]$ to be packed and the desired tile tensor shape: $T_A=pack(A,[\frac{n_1}{t_1},\ldots,\frac{n_k}{t_k}])$. The \emph{pack} operator computes the external tensor using the tiling process described above, and stores along-side it the tile tensor shape, to form the full tile tensor $T_A$. We can retrieve $A$ back using the {\em unpack} operation: $A=unpack(T_A)$.
As with regular tensors, we sometimes refer to a tile tensor $T_A$ together with its shape: $T_A[\frac{n_1}{t_1},\ldots,\frac{n_k}{t_k}]$.

\subsection{Replication}

\begin{figure}[htp]
 
  \begin{subfigure}[t]{0.22\linewidth}
    \centering
    \includegraphics[scale=0.3]{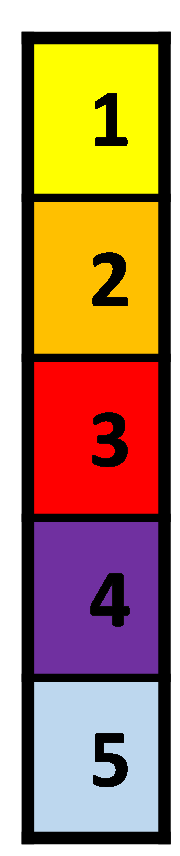}
    \caption{$V[5,1]$}
    \label{fig:V5_1}
  \end{subfigure}%
  \hfill
  \begin{subfigure}[t]{0.25\linewidth}
    \centering
    \includegraphics[scale=0.3]{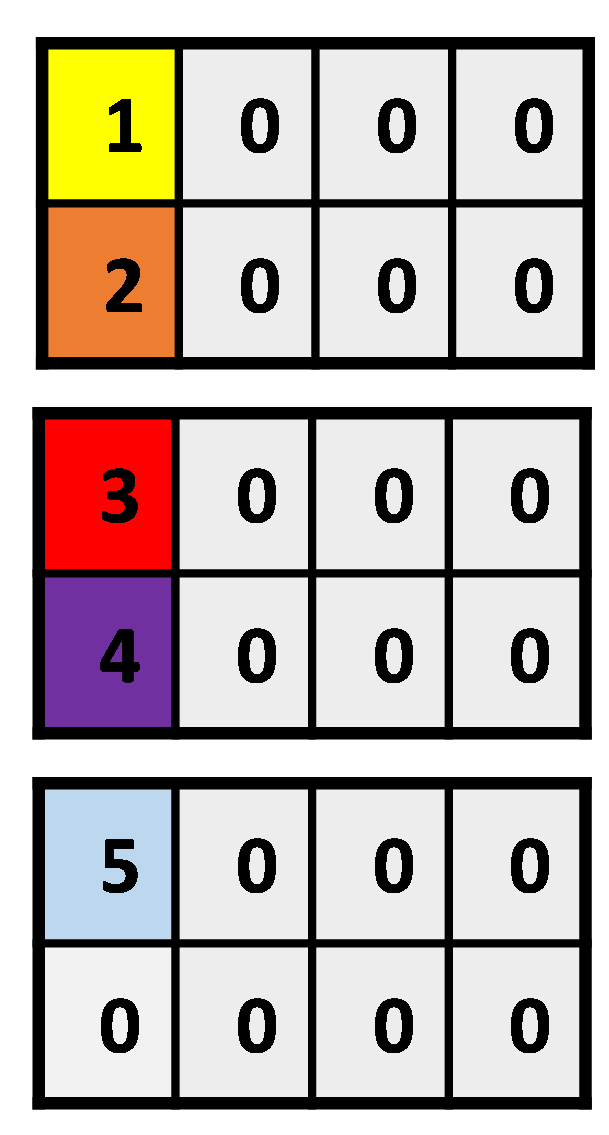}
    \caption{$T_V[\frac{5}{2},\frac{1}{4}]$}
    \label{fig:V5_2__1_4}
  \end{subfigure}%
  \hfill
  \begin{subfigure}[t]{0.25\linewidth}
    \centering
    \includegraphics[scale=0.3]{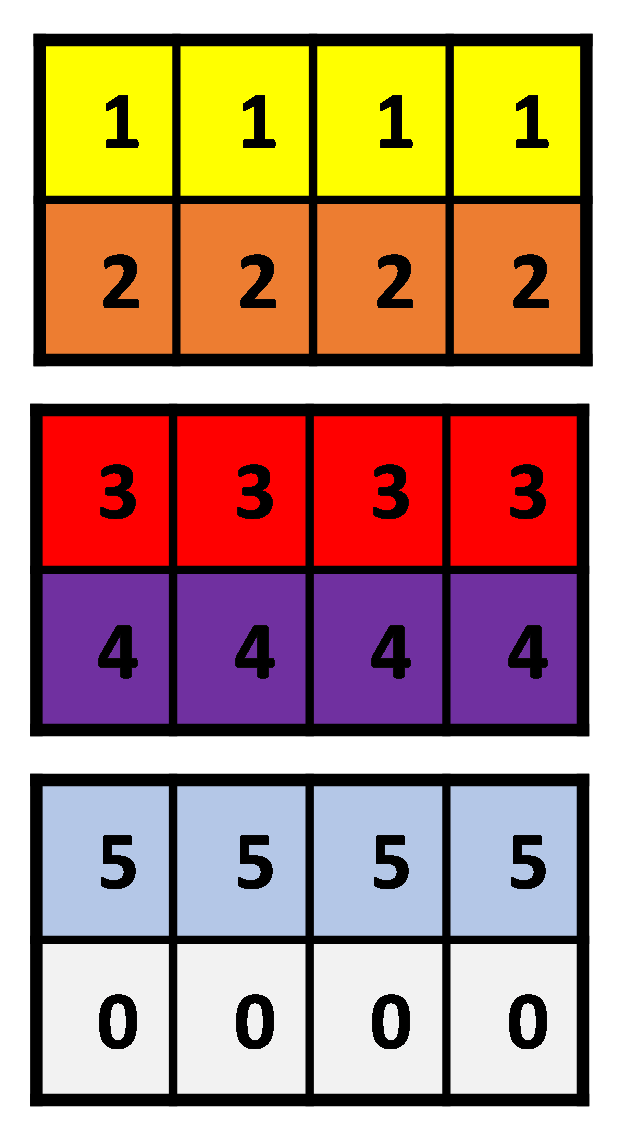}
    \caption{$T_V'[\frac{5}{2},\frac{*}{4}]$}
    \label{fig:V5_2__S_4}
  \end{subfigure}%
  \hfill
  \begin{subfigure}[t]{0.25\linewidth}
    \centering
    \includegraphics[scale=0.3]{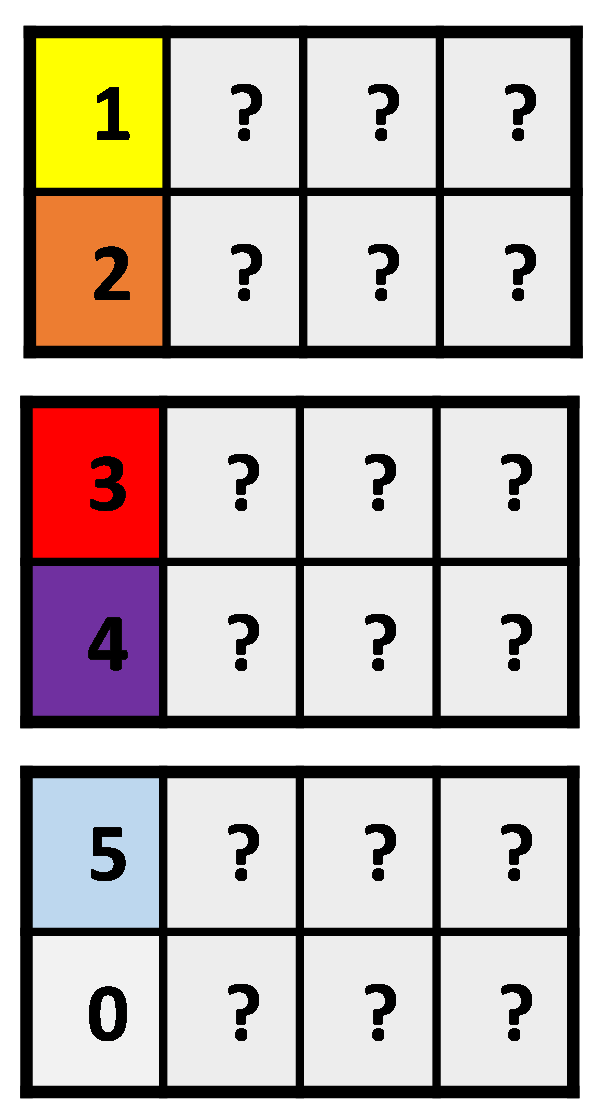}
    \caption{$T_V''[\frac{5}{2},\frac{1?}{4}]$}
    \label{fig:V5_2__1Q_4}
  \end{subfigure}%

  \caption{Packing $V[5,1]$ into tile tensors using different tile tensor shapes. Every rectangle represents a tile. Panel (c) demonstrates tiling with replication, where the packing process computes $V'=broadcast(V,[5,4])$ and tiles $V'$. Panel (d) demonstrates unknown values along the second dimension using question mark symbols.}
  \label{fig:VV5}
\end{figure}
\raggedbottom

For some computations it is useful to have the tensor data replicated several times inside the tile slots.
The tile tensor shape indicates this by using the $\frac{*}{t_i}$ notation. It implies that $n_i=1$, but each element of the original tensor is replicated $t_i$ times along the $i$'th dimension. When packing a tensor $A[n_1,\ldots,n_k]$ and $n_i=1$, and with a tile tensor shape specifying $\frac{*}{t_i}$,  then the packing operation performs $broadcast(A,[n_1,\ldots,t_i,\ldots,n_k])$ and tiles the result. The unpacking process shrinks the tensor back to its original size. The replications can either be ignored, or an average of them can be taken; this is useful if the data is stored in a noisy storage medium, as in approximate \gls{HE} schemes. Figure~\ref{fig:VV5} demonstrates packing $V[5,1]$ with different tile tensor shapes. Note that we only replicate inside a tile. When broadcasting requires replicating whole tiles it is possible to hold only one copy of it and use it multiple times.

\subsection{Unknown Values}

When tensors are packed into tile tensors, unused slots are filled with zeroes, as shown in Figure~\ref{fig:MM56}.  Section~\ref{subsec:intro_ops} shows how to apply operators on tile tensors, and then unused slots might get filled with arbitrary values. Although these unused slots are ignored when the tile tensor is unpacked, the presence of arbitrary values in them can still impact the validity or performance of applying additional operators. To reflect this state, the tile tensor shape contains an additional flag per dimension, denoted by the symbol ``?'', indicating the presence of unknown values.

Figure~\ref{fig:V5_2__1Q_4} shows a tile tensor with the shape  $[\frac{5}{2},\frac{1?}{4}]$. The ``$?$'' in the second dimension indicates that whenever we exceed the valid range of the packed tensor along this dimension, we may encounter arbitrary unknown values. However, it still holds that $V=unpack(T_V)$, as these unused slots are ignored.

\subsection{Operators}
\label{subsec:intro_ops}
Tile tensor operators are homomorphic operations between tile tensors and the packed tensors they contain.
For two tile tensors $T_A$ and $T_B$, and a binary operator $\odot$, it holds that $unpack(T_A \odot T_B)=unpack(T_A)\odot unpack(T_B)$. Unary operators are similarly defined. 

\begin{remark}
The above operations involve two levels of homomorphism. One homomorphism is between direct operations on ordinary tensors $A$, $B$ and operating on the tile tensors $T_A$, $T_B$ that contain them. The tile tensor operators are implemented by performing operations on encrypted tiles. Due to the properties of homomorphic encryption (see Section~\ref{subsection:background_he}), these are homomorphic to ordinary operations on the plaintext content of the tiles.
\end{remark}

Binary elementwise operators  are implemented by applying the operator on the external tensors tile-wise, and the tile tensor shape is updated to reflect the shape of the result. If the inputs have identical shapes then so do the results, e.g.,
$T_M''[\frac{5}{2},\frac{6}{4}]$ of Figure~\ref{fig:M5_2__6_4} can be multiplied with an identically packed matrix $T_N[\frac{5}{2},\frac{6}{4}]$, resulting in $T_R[\frac{5}{2},\frac{6}{4}]$, where $R=M*N$. As with regular tensors, the tile tensor shapes need not be identical, but compatible. 
Compatible tile tensor shapes have the same number of dimensions, and for each dimension specification they are either identical, or one is $\frac{*}{t_i}$ and the other is $\frac{n_i}{t_i}$. 
The intuition is that if the tensor is already broadcast inside the tile, it can be further broadcast to match any size by replicating the tile itself.
For example, for $T_V'[\frac{5}{2},\frac{*}{4}]$ of Figure~\ref{fig:V5_2__S_4} we can compute $T_M''*T_V'$ resulting in $T_R'=[\frac{5}{2},\frac{6}{4}]$. We can also compute $T_M''+T_V''$, but this results in $T_R''[\frac{5}{2},\frac{6?}{4}]$, i.e., with unknown values in unused slots along the second dimension. This occurs because in $T_V'$ this dimension is filled with replicated values, and after the addition they fill the unused slots of the result. Computing $T_M''*T_V$ (for $T_V$ from Figure~\ref{fig:V5_2__1_4}) is illegal because their shapes are not compatible.
For the full set of rules for elementwise operators see Appendix~\ref{appendix:tt_def}.

The $sum$ operator is also defined homomorphically: \\
$unpack(sum(T_A,i)) = sum(unpack(T_A),i)$. It works by summing over the external tensor along the $i$'th dimension, then by summing inside each tile along the $i$'th dimension. 
In an \gls{HE} environment, the latter summation requires using the rotate-and-sum algorithm \cite{TILETENSORS}.
Generally, the sum operator reduces the  $i$'th dimension and the resulting tile tensor shape changes to $\frac{1?}{t_i}$. However, there are some useful special cases. If $t_i=1$, then it is reduced to $\frac{1}{1}$ or simply $1$. When $i$ is the smallest $i$ such that $t_i>1$, the dimension reduces to  $\frac{*}{t_i}$, i.e., the sum results are replicated. This is due to properties of the rotate-and-sum algorithms. It is a useful property, since this replication is sometimes needed for compatibility with another tile tensor. For example,
let $T_A$ be a tile tensor with the shape $[4,\frac{3}{8},\frac{5}{16}]$.
Then, $sum(T_A,1)$ is of shape $[1,\frac{3}{8},\frac{5}{16}]$;
$sum(T_A,2)$ is of shape $[4,\frac{*}{8},\frac{5}{16}]$ and
$sum(T_A,3)$ is of shape $[4,\frac{3}{8},\frac{1?}{16}]$.

Three other operators used in this paper do not change the packed tensor, just the external tensor and tile tensor shape. The $clear(T_A)$ operator clears unknown values by multiplying with a mask containing ones for all used slots, i.e., it removes the "?" from the tile tensor shape. For example, $clear(T_V''[\frac{5}{2},\frac{1?}{4}])=T_V[\frac{5}{2},\frac{1}{4}]$ (see Figure~\ref{fig:VV5}). The $rep(T_A,i)$ operator assumes the $i$'th dimension is $\frac{1}{t_i}$, and replicates it to $\frac{*}{t_i}$, using a rotate-and-sum algorithm.
The $flatten(T_A,i,j)$ operator flattens dimensions $i$ through $j$ assuming they are all replicated. This is done trivially by just changing the meta data, e.g., $flatten(T_A[\frac{3}{4},\frac{*}{8},\frac{*}{2},\frac{5}{32}],2,3)$ results with $T_A'[\frac{3}{4},\frac{*}{16},\frac{5}{32}]$.

\subsection{Higher Level Operators}
\label{subsection:high_level_ops}
Using elementwise operators and summation, we can perform various algebraic operations on tile tensors.

{\bf Matrix-vector multiplication.} Given a matrix $M[a,b]$ and a vector $V[b]$, we reshape $V$ to $V[1,b]$ for compatibility, and pack both tensors into tile tensors as $T_M[\frac{a}{t_1},\frac{b}{t_2}]$, and $T_V[\frac{*}{t_1},\frac{b}{t_2}]$, for some chosen tile shape $[t_1,t_2]$. We can multiply them using:
\begin{equation}
\label{equ_ttmatvec1}
T_R[\frac{a}{t_1},\frac{1?}{t_2}]=sum(T_M[\frac{a}{t_1},\frac{b}{t_2}]*T_V[\frac{*}{t_1},\frac{b}{t_2}],2).
\end{equation}

The above formula works for any value of $a, b, t_1, t_2$. This is because the tile tensor shapes of  $T_M$ and $T_V$ are compatible, and therefore, due to the homomorphism, this computes $R[a,1]=sum(M[a,b]*V[1,b],2)$, which produces the correct result as explained in Section~\ref{subsection:tensor_basics}.

A second option is to initially transpose both $M$ and $V$ and pack them in tile tensors $T_M[\frac{b}{t_1},\frac{a}{t_2}]$ and $T_V[\frac{b}{t_1},\frac{*}{t_2}]$. Now we can multiply them as:
\begin{equation}
\label{equ_ttmatvec2}
T_R[\frac{*}{t_1},\frac{a}{t_2}]=sum(T_M[\frac{b}{t_1},\frac{a}{t_2}]*T_V[\frac{b}{t_1},\frac{*}{t_2}],1).
\end{equation}

This computes the correct result using the same reasoning as before. The benefit here is that the result $T_R[\frac{*}{t_1},\frac{a}{t_2}]$ is replicated along the first dimension due to the properties of the sum operator (Section~\ref{subsec:intro_ops}). Thus, it is ready to play the role of $T_V$ in  Formula~\ref{equ_ttmatvec1}, and we can perform two matrix-vector multiplications consecutively without any processing in between. The output of Formula~\ref{equ_ttmatvec1} can be processed to fit as input for Formula~\ref{equ_ttmatvec2} using $rep(clean(T_R),2)$.

{\bf Matrix-matrix multiplication.} The above reasoning easily extends to matrix-matrix multiplication as follows. Given matrices $M_1[a,b]$ and $M_2[b,c]$, we can compute their product using either of the next two formulas, where in the second one we transpose $M_1$ prior to packing. As before, the result of the second fits as input to the first.

\begin{equation}
T_R[\frac{a}{t_1},\frac{1?}{t_2},\frac{c}{t_3}]=sum(T_{M_1}[\frac{a}{t_1},\frac{b}{t_2},\frac{*}{t_3}]*T_{M_2}[\frac{*}{t_1},\frac{b}{t_2},\frac{c}{t_3}],2).
\label{equ:matmul1}
\end{equation}

\begin{equation}
\label{equ_ttmatmat2}
T_R[\frac{*}{t_1},\frac{a}{t_2},\frac{c}{t_3}]=sum(T_{M_1}[\frac{b}{t_1},\frac{a}{t_2},\frac{*}{t_3}]*T_{M_2}[\frac{b}{t_1},\frac{*}{t_2},\frac{c}{t_3}],1).
\end{equation}

\begin{example}
The product $R[100,60]=\prod_{i=1}^{4}{M_i}$ of the four matrices $M_1[100,90]$, $M_2[90,80]$, $M_3[80,70]$, and $M_4[70,60]$ is computed by packing the matrices in tile tensors 
$T_{M_1}[\frac{90}{t_1},\frac{100}{t_2},\frac{*}{t_3}]$, $T_{M_2}[\frac{90}{t_1},\frac{80}{t_2},\frac{*}{t_3}]$, $T_{M_3}[\frac{70}{t_1},\frac{80}{t_2},\frac{*}{t_3}]$, and $T_{M_4}[\frac{70}{t_1},\frac{*}{t_2},\frac{60}{t_3}]$
and computing
\begin{align*}
T_{X_1}[\frac{90}{t_1},\frac{1?}{t_2},\frac{60}{t_3}]
  &= sum\left(T_{M_2}* 
    \left(sum(T_{M_3}*T_{M_4},1\right)
  \right),2)\\
T_R[\frac{*}{t_1},\frac{100}{t_2},\frac{60}{t_3}] &= 
sum\left(
  T_{M_1} * \left(
    rep\left(clean\left(T_{X_1}\right),2\right)
    \right),1
  \right)
\end{align*}
\end{example}

\subsection{Interleaved Tiling}
\label{subsection:interleaved}
Another option for tiling is denoted by the symbol ``$\sim$'' in the tile tensor shape. This symbol indicates that the tiles do not cover a contiguous block of the tensor, but are spread out in equal strides.
If the dimensions are interleaved, an element of the original tensor $A(c_1,c_2,\ldots,c_k)$ will be mapped to tile indices $a_i=c_i\mod e_i$, and indices inside the tile $b_i=\floor{\frac{c_i}{e_i}}$ (where $e_i$ is the size of the external tensor, see Section~\ref{subsec:tiling}). See Figures~\ref{fig:conv1a} and~\ref{fig:conv1b} for an example. 

For each dimension, we can specify separately whether it is interleaved or not. For example, in $[\frac{5}{2},\frac{6\sim}{4}]$ only the second dimension is interleaved. Also, although with basic tiling it holds that $e_i=\ceil{\frac{n_i}{t_i}}$, for interleaved tiling it is sometimes useful to have larger values for $e_i$. In this case, this value can be explicitly stated using the notation: $\frac{n_i\sim e_i}{t_i}$.

Interleaved dimensions fit seamlessly with all the operators described above, and are useful for computing convolutions. See Section~\ref{section:convolution} for more details.

\section{Convolution Using Tile Tensors}
\label{section:convolution}
In this section we describe our novel method to compute convolution. Compared to previous work (e.g. \cite{GALA, GAZELLE2018}) our method has two advantages: (i) it is more efficient when the input is a large image and (ii) it allows  efficient computation of consecutive convolution layers in an \gls{HE} only system.

We first consider in Section~\ref{subsec:conv_interleaved} the convolution problem in its simplest form: a single, one-channel image, and a single filter $F[w_F,h_F]$. We extend it to convolution with strides and multiple channels, filters, and batching in Section~\ref{subsec:conv_channels_filter}.

\subsection{Convolution with Interleaved Dimensions}
\label{subsec:conv_interleaved}

We now show how interleaved dimensions (see Section~\ref{subsection:interleaved}) can be used to efficiently compute convolution.

\begin{figure*}[ht]
    \centering
    \begin{subfigure}{0.31\linewidth}
        \centering
        \includegraphics[scale=0.25]{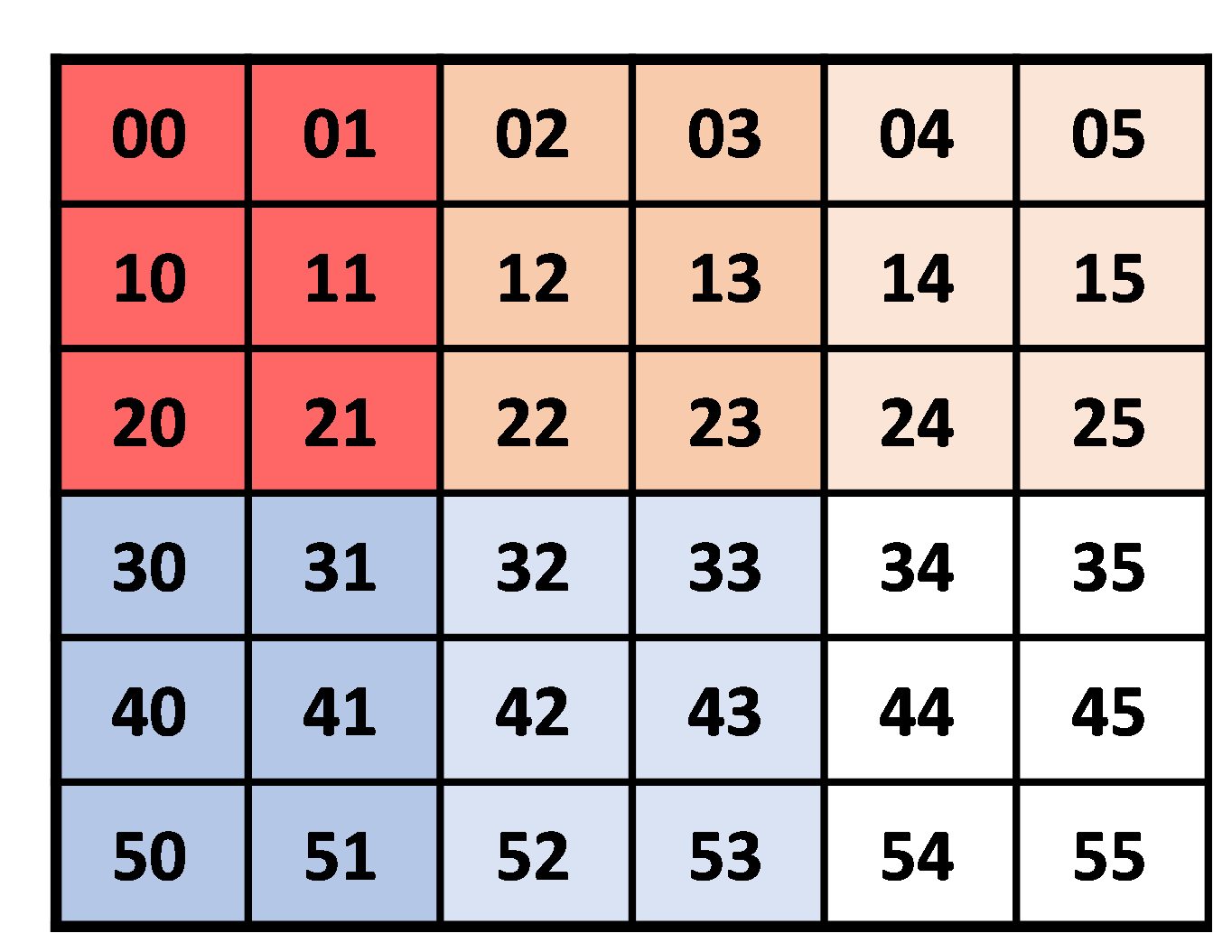}
        \caption{$M[6,6]$}
        \label{fig:conv1a}
  \end{subfigure}
  \hfill
  \unskip {\vrule width 1pt}
  \begin{subfigure}{0.33\linewidth}
        \centering
        \includegraphics[scale=0.25]{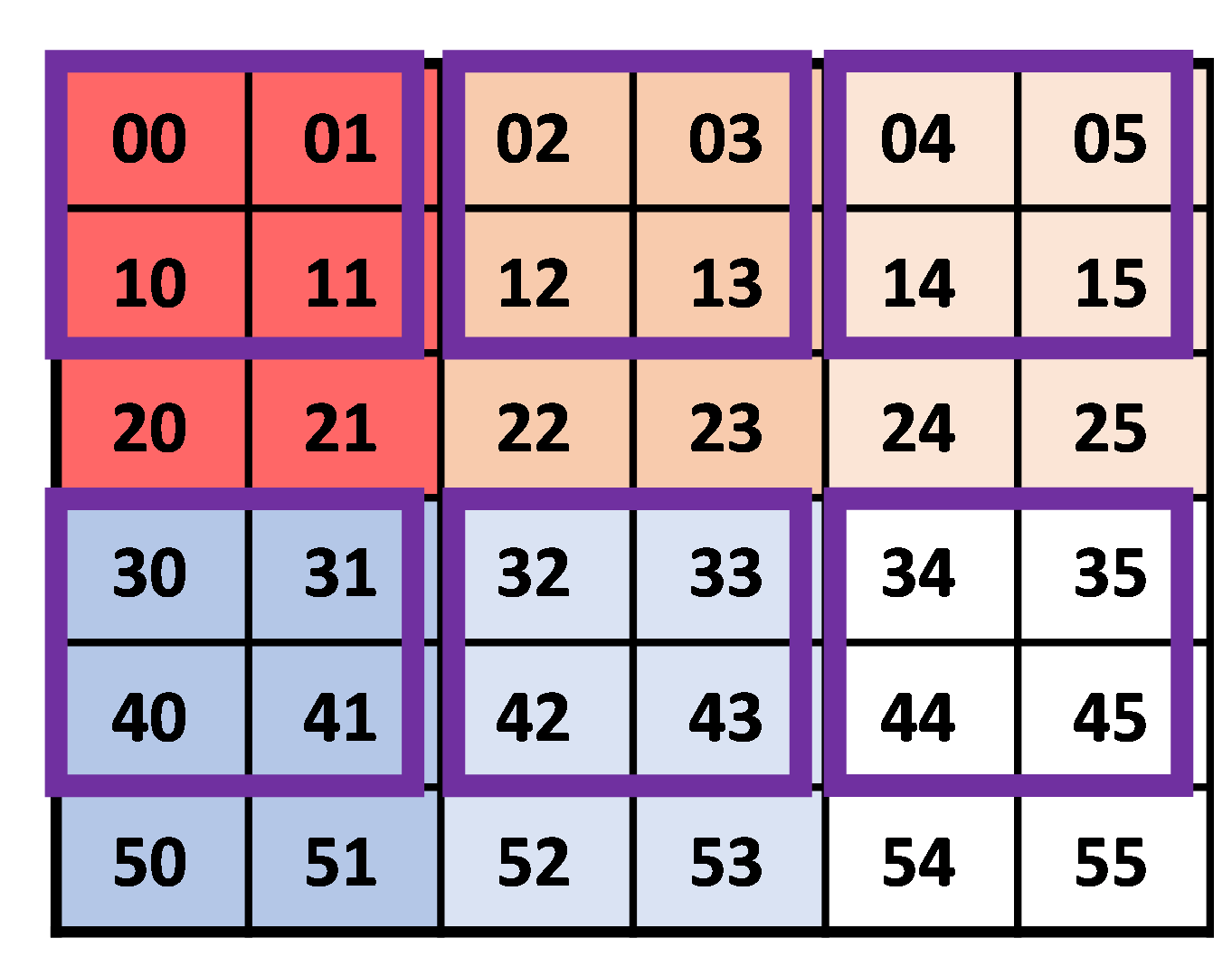}
        \vspace{16pt}
  \end{subfigure}
  \hfill
  \unskip {\vrule width 1pt}
  \begin{subfigure}{0.33\linewidth}
        \centering
        \includegraphics[scale=0.25]{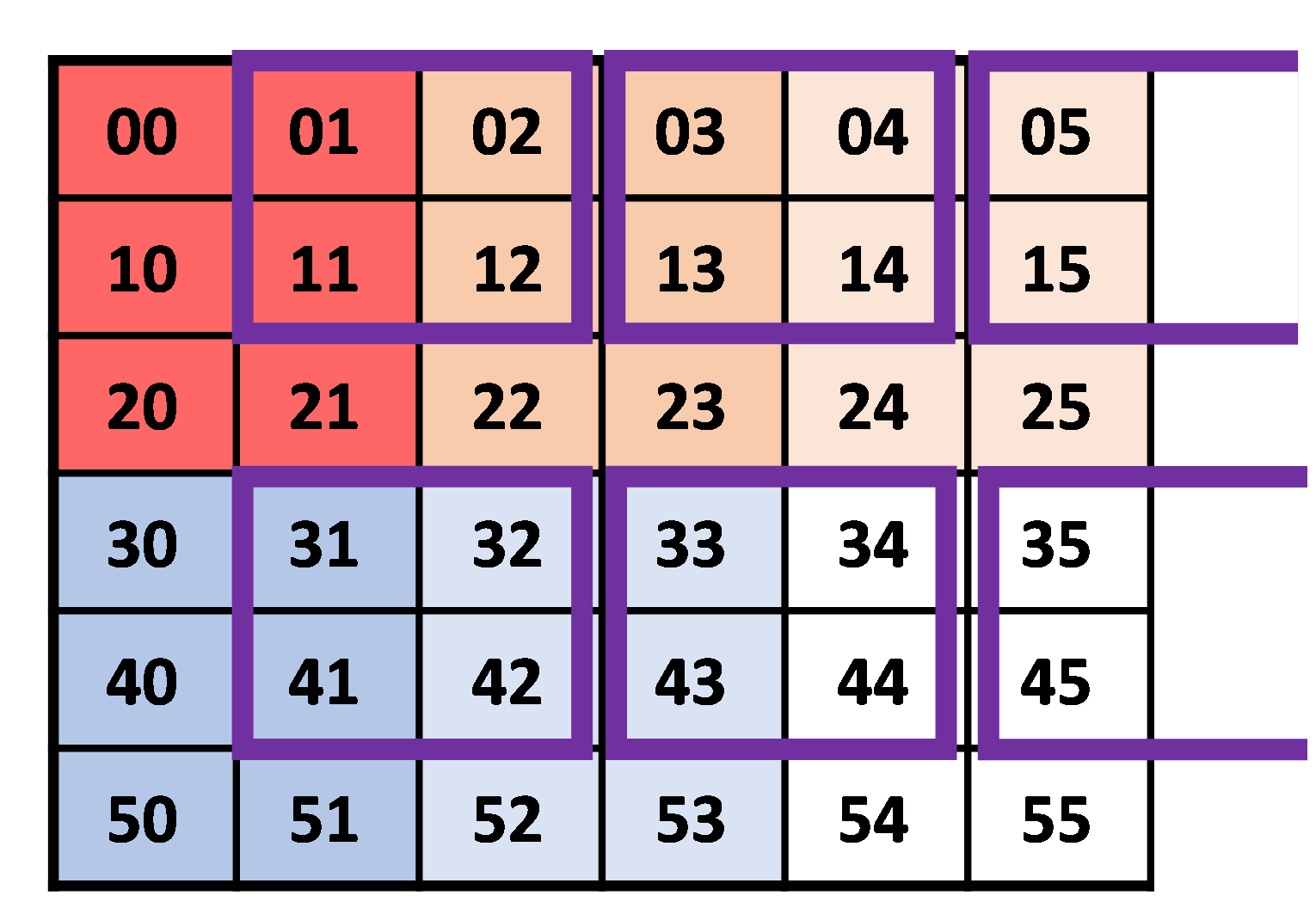}
        \vspace{16pt}
  \end{subfigure}

    \begin{subfigure}{0.31\linewidth}
        \centering
        \includegraphics[scale=0.25]{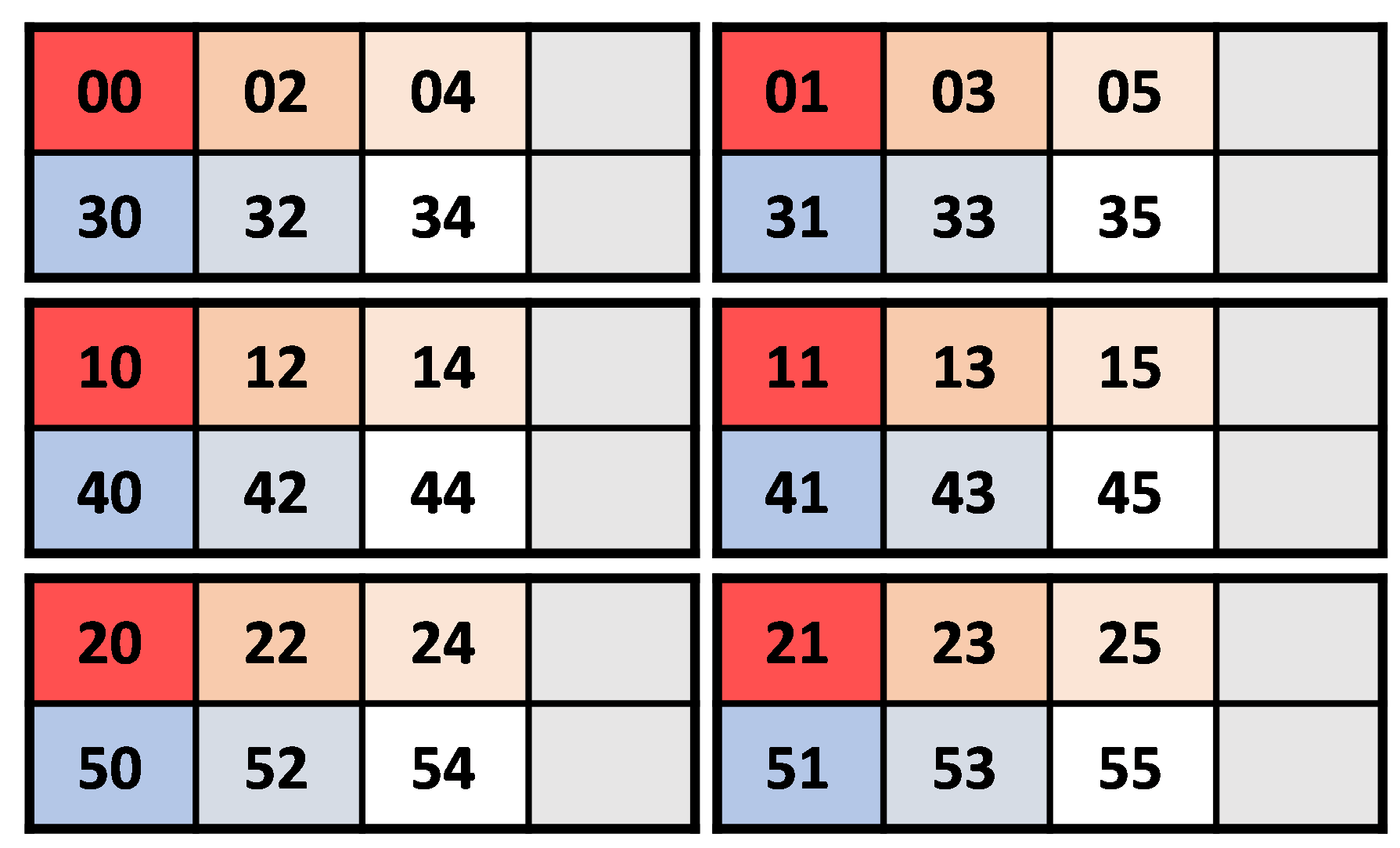}
    \caption{$T_M[\frac{~6\sim}{2},\frac{~6\sim}{4}]$}
    \label{fig:conv1b}
  \end{subfigure}
  \hfill
  \unskip {\vrule width 1pt}
  \begin{subfigure}{0.33\linewidth}
        \centering
        \vspace{25pt}
        \includegraphics[scale=0.25]{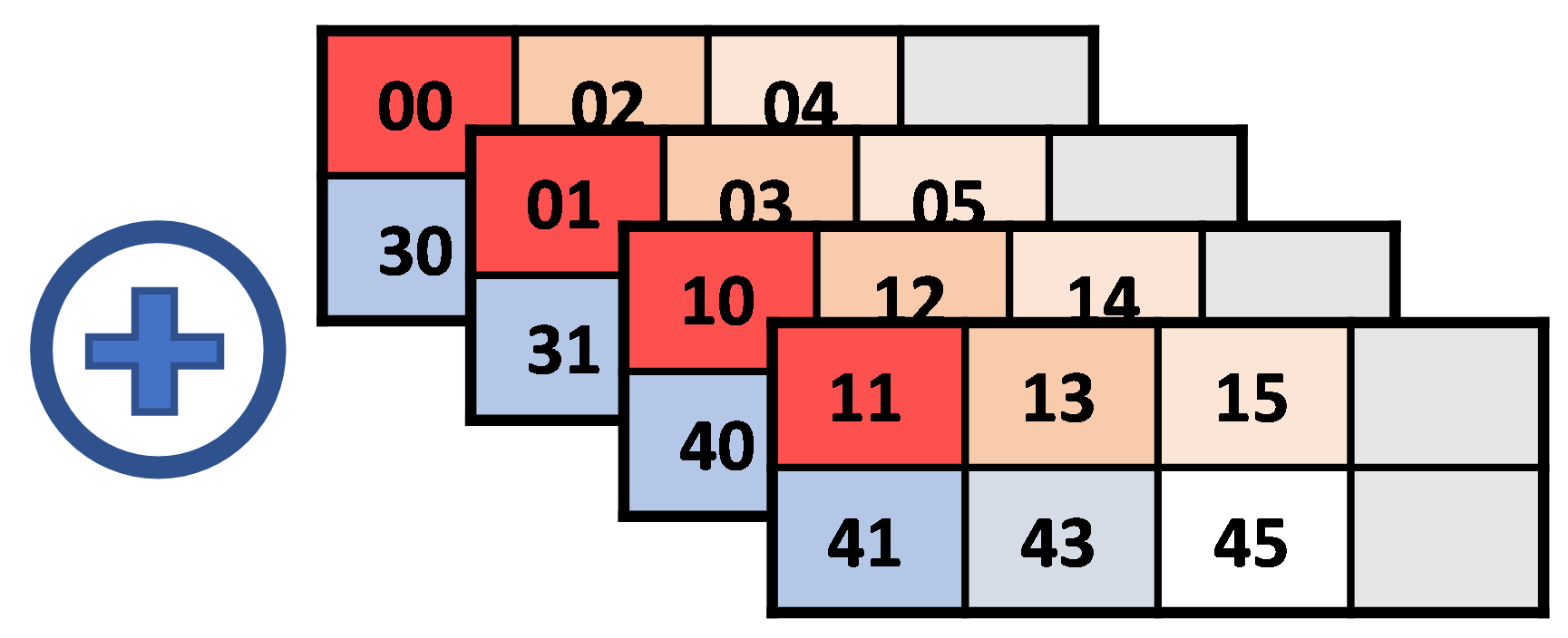}
        \vspace{17pt}
        \caption{{First convolution iteration on six parallel windows, ignoring the extra two slots.}}
        \label{fig:conv1c}
  \end{subfigure}
  \hfill
  \unskip {\vrule width 1pt}
  \begin{subfigure}{0.33\linewidth}
        \centering
        \vspace{27pt}
        \includegraphics[scale=0.25]{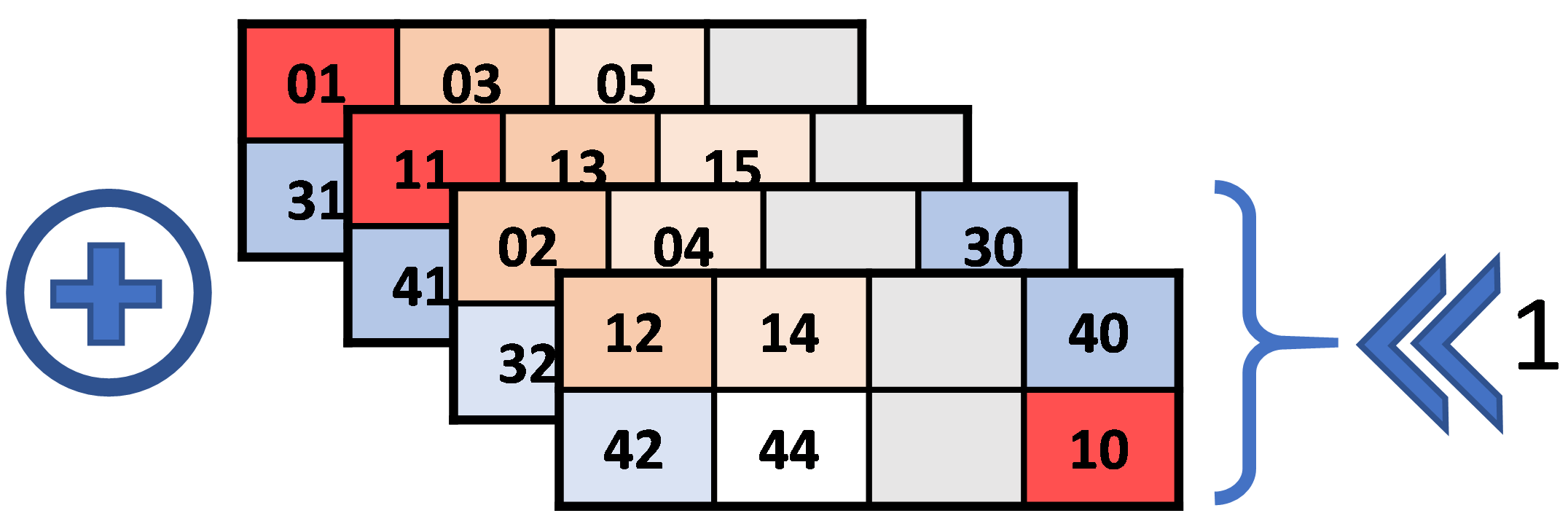}
        \vspace{1pt}
        \caption{{Second convolution iteration on four parallel windows, ignoring the extra four slots.}}
        \label{fig:conv1d}
  \end{subfigure}
    \caption{Packing a matrix $M[6,6]$ into  $T_M[\frac{6\sim}{2},\frac{6\sim}{4}]$ and performing two 8-parallel kernel evaluations using a $2\times2$ filter (purple). The upper figures illustrate the convolution operator on the $M[6,6]$ matrix. The lower figures illustrate the same operation using tiles representation. 
    Here, $\oplus$ denotes component-wise summation of tiles and $\ll 1$ denotes left circular rotation by $1$.}
  \label{fig:conv1}
\end{figure*}

Figures~\ref{fig:conv1a} and~\ref{fig:conv1b} show a matrix $M[6,6]$ packed in the tile tensor $T_M[\frac{6\sim}{2},\frac{6\sim}{4}]$. Here, the tile shape is $[2, 4]$ and the external tensor shape is $[3, 2]$. Every tile contains a $2 \times 4$ sub-matrix, but instead of being contiguous, it is a set of elements spaced evenly in the matrix.
We use the same color to denote elements that are mapped to the same slot in different tiles. For example, the elements $00, 01, 10, 11, 20, 21$ are all colored red as they are mapped to slot $(0,0)$ in 6 different tiles. This coloring divides the original matrix into contiguous regions.

The interleaved packing allows for a more efficient implementation of Equation~\ref{equ:conv_one} with respect to runtime and storage.
Intuitively, we use the SIMD to compute multiple elements of the output in a single operation.
The filter is packed simply as $T_F[w_F,h_F, \frac{*}{t_1},\frac{*}{t_2}]$. I.e., it has $w_F h_F$ tiles, each containing one value of the filter in all slots. This allows multiplying each image tile with each value of the filter.

For example, Figure~\ref{fig:conv1c} shows a computation of the convolution output when the filter is placed at the top left position. The SIMD nature of the computation computes the output in other regions as well. The result is a single tile, where each slot contains the convolution result of the corresponding region, such that this tile is packed in the same interleaved packing scheme as the input tiles.

A more complicated example is given in Figure~\ref{fig:conv1d}. Here the filter is placed one pixel to the right. As a result, the filter needs to be multiplied by elements that appear in different regions, i.e. they are mapped to slots of different indices.
In this case we need to rotate the tiles appropriately. For example, placing the filter with its upper left corner on pixel $(0,1)$, the convolution is computed using the $(0,0)$ slot of tiles $(0,1)$ and $(1,1)$ and slot $(0,1)$ of tiles $(0,0)$ and $(1,0)$. The latter two are therefore rotated to move the required value to slot $(0,0)$ as well.

The total cost of convolution when using this packing is summarized in the following lemma.

\begin{restatable}{lemma}{convlemma}
\label{lem:conv1}
Let $s$ be the number of slots in a ciphertext. Then, given an input image $I[w_I,h_I]$ and a filter $F[w_F,h_F]$, packing $I$ as\\ $T_I[\frac{w_I\sim}{t_1},\frac{h_I\sim}{t_2}]$ and the filter as $T_F[w_F,h_F, \frac{*}{t_1},\frac{*}{t_2}]$, 
convolution can be computed using:
$\BigO(\ceil {w_I h_I w_F h_F / s}) \text{ multiplications, and}$
$\BigO(w_F \ceil{\frac{w_I}{t_1}} + h_F \ceil{\frac{h_I}{t_2}} + w_F h_F) \text{ rotations}.$
The input is encoded in 
$\BigO(w_I h_I / s)$ ciphertexts.
\end{restatable}

\begin{proof}
See Appendix \ref{sec:proof}.
\end{proof}

The output of the convolution is a tile tensor $T_O[\frac{w_O\sim?}{t_1},\frac{h_O\sim?}{t_2}]$. The unknown values are introduced by filter positions that extend beyond the image, as shown in Figure~\ref{fig:conv1d}. Note further that the external sizes $e_1=\ceil{\frac{w_I}{t_1}}$ and $e_2=\ceil{\frac{h_I}{t_2}}$ of the tile tensor $T_I$ remain the same in $T_O$, and they may be larger than those actually required to hold the tensor $O[w_O,h_O]$. Hence, a more accurate depiction of $T_O$'s shape is $T_O[\frac{w_O\sim e_1?}{t_1},\frac{h_O\sim e_2?}{t_2}]$, but we will ignore this technicality from here on. 

\subsection{Handling Strides, Batching and Multiple Channels and Filters}
\label{subsec:conv_channels_filter}
In this section we extend the simple description given in Section~\ref{subsec:conv_interleaved}. We first show how our convolution algorithm extends to handle multiple channels, multiple filters, and batching. We then show how we handle striding.

Let the input be a tensor of images $I[w_I,h_I,c,b]$, where $c$ is the number of channels and $b$ is the batch size. Then we pack it as
$T_I[\frac{w_I\sim}{t_1},\frac{h_I\sim}{t_2},\frac{c}{t_3},\frac{b}{t_4},\frac{*}{t_5}]$.
Also, we pack the filters $F[w_F,h_F,c,f]$, where $f$ is the number of filters, as $T_F[w_F,h_F,\frac{*}{t_1},\frac{*}{t_2},\frac{c}{t_3},\frac{*}{t_4},\frac{f}{t_5}]$, where $t_i \in \mathcal{N}$ and $\prod t_i = s$.

The convolution is computed similarly to the description in Section~\ref{subsec:conv_interleaved},
multiplying tiles of $T_I$ with the appropriate tiles
of $T_F$.
The result is a tile tensor of shape $T_O[\frac{w_O\sim?}{t_1},\frac{h_O\sim?}{t_2},\frac{c}{t_3},\frac{b}{t_4},\frac{f}{t_5}]$. Summing over the channel (i.e., third) dimension using $\BigO(\ceil{\frac{w_O}{t_1}}\ceil{\frac{h_O}{t_2}}\ceil{\frac{b}{t_4}}\ceil{\frac{f}{t_5}}$ $\log t_3)$ rotations, we obtain $T_O[\frac{w_O\sim?}{t_1},\frac{h_O\sim?}{t_2},\frac{1?}{t_3},\frac{b}{t_4},\frac{f}{t_5}]$, where the $\log t_3)$ factor comes from the sum-and-rotate algorithm \cite{TILETENSORS}.

For bigger strides, $\strideh>1$ (resp. $\stridew > 1$), we require that either $t_1 = 1$ (resp. $t_2=1$) or $\ceil{\frac{h_I}{t_2}} \mod \strideh = 0$ (resp. $\ceil{\frac{w_I}{t_1}} \mod \stridew=0$). Then, our implementation trivially skips $\stridew$ ciphertexts in every row and $\strideh$ ciphertexts in every column.

\vspace{-5pt}
\subsection{A Sequence of Convolutions}
\label{subsec:conv_sequence}
In this section we discuss how to implement a sequence of multiple convolution layers. This is something that is common in neural networks.
One of the advantages of our tile tensor method is that the output of one convolution layer can be easily adjusted to be the input of the next convolution layer.

Assume we are given an input batch tensor
$I[w_I,h_I,c,b]$ and a sequence of convolution layers with the $l$'th layer having a filter tensor $F^l[w_F^l,h_F^l,c^l,f^l]$. For the first layer we have $c^1=c$, and for $l>1$ we have $c^l=f^{l-1}$.
As before, we pack the input tensor as $T_I[\frac{w_I\sim}{t_1},\frac{h_I\sim}{t_2},\frac{c}{t_3},\frac{b}{t_4},\frac{*}{t_5}]$.
For odd layers, $l = 2\ell +1$, we pack the filter tensor as before $T_F^l[w_F^l,h_F^l,\frac{*}{t_1},\frac{*}{t_2},\frac{c}{t_3},\frac{*}{t_4},\frac{f^l}{t_5}]$.
The output is then $T_O[\frac{w_O^l\sim ?}{t_1},\frac{h_O^l\sim ?}{t_2},\frac{1?}{t_3},\frac{b}{t_4},\frac{f^l}{t_5}]$.
For even layers, $l = 2\ell$, we introduce this packing for the filters:
$T_F^l[w_F^l,h_F^l,\frac{*}{t_1},\frac{*}{t_2},\frac{f}{t_3},\frac{*}{t_4},\frac{c}{t_5}]$.

As can be seen, the shapes of the layer outputs do not match the shapes of the inputs of the subsequent layers. We now show how to solve it and thus allow for a sequence of convolution layers.

To make an output of an odd layer suitable for the next even layer, we clear the unknowns by multiplying with a mask and then replicate the channel dimension. We then get a tile tensor of this shape:
$T_O[\frac{w_O^l\sim ?}{t_1},\frac{h_O^l\sim ?}{t_2},\frac{*}{t_3},\frac{b}{t_4},\frac{f^l}{t_5}]$,
which matches the input format of the next layer since $f^l=c^{l+1}$. To make an output of an even layer suitable for the next odd layer,
we similarly clean and replicate along the filter dimension.

We note that changing the order of the dimensions leads to a small improvement.
The improvement comes because summing over the first dimension ends up with a replication over this dimension.
Therefore, setting the channel dimension as the first dimension saves us the replication step when preparing the input to an even layer. We can skip cleaning as well, since the unknown values along the image width and height dimensions do no harm.
Alternatively, the filter dimension can be set as first and then the replication step can be skipped when preparing the input for an odd layer.

\subsection{Na\"ive Convolution Methods }
\label{sec:naive_conv}

The above method reduces to a simple method known by various names such as SIMD packing when $t_1=t_2=t_3=t_5=1$. In this case, every element in the tensors for the images and filters is stored in a separate ciphertext, and the slots are only used for batching.
In this paper, we further use the reduction to matrix multiplication as described in Section~\ref{subsec:background_im2col}.
It is applicable only for \glspl{NN} with one convolutional layer.

\section{The Optimizer}\label{section:optimizer}

The packing optimizer is responsible for finding the most efficient packing arrangement for a given computation, as well as the optimal configuration of the underlying \gls{HE} library. This relieves the users from the need to handle these \gls{HE} related complexities. The users only need to supply the model architecture e.g., a \gls{NN} architecture, in some standard file format. The optimizer automatically converts it to an \gls{HE} computation with optimal packing and optimal \gls{HE} library configuration. Users can further supply constraints such as the required security level or maximal memory usage, and choose an optimization target, whether to optimize for CPU time, latency or throughput, or optimize for memory usage. 
Our optimizer has many similarities to the one of CHET. However, it has one principal difference, it operates over and understands our concept of tile tensors. This allows it, for example, to explore different intermediate batching levels and to suggest different tradeoffs between latency and amortized latency, which was not discussed or analyzed in CHET.

The optimizer needs to choose among different possible configurations of the \gls{HE} library, as well as different packing techniques to support certain operators (see Section~\ref{section:convolution}). It also chooses the tile shape, i.e., the values of $t_1,t_2,\ldots$, in the tile tensor shapes. For example, consider an \gls{HE} scheme configured to have  $16{,}384$ slots in each ciphertext. Since our convolution operator uses five-dimensional tiles, the number of possible tuples $t_1,\ldots,t_5$ such that $\prod_i t_i=16{,}384$ is $\binom{\log_2(16{,}384)+5-1}{5-1}=3{,}060$. We use the term \emph{configuration} to refer to a complete set of options the optimizer can choose (\gls{HE} configuration parameters, tile shape, and other packing options).

\begin{figure}[ht!]
    \centering
    \vspace{-10pt}
    \includegraphics[width=\linewidth]{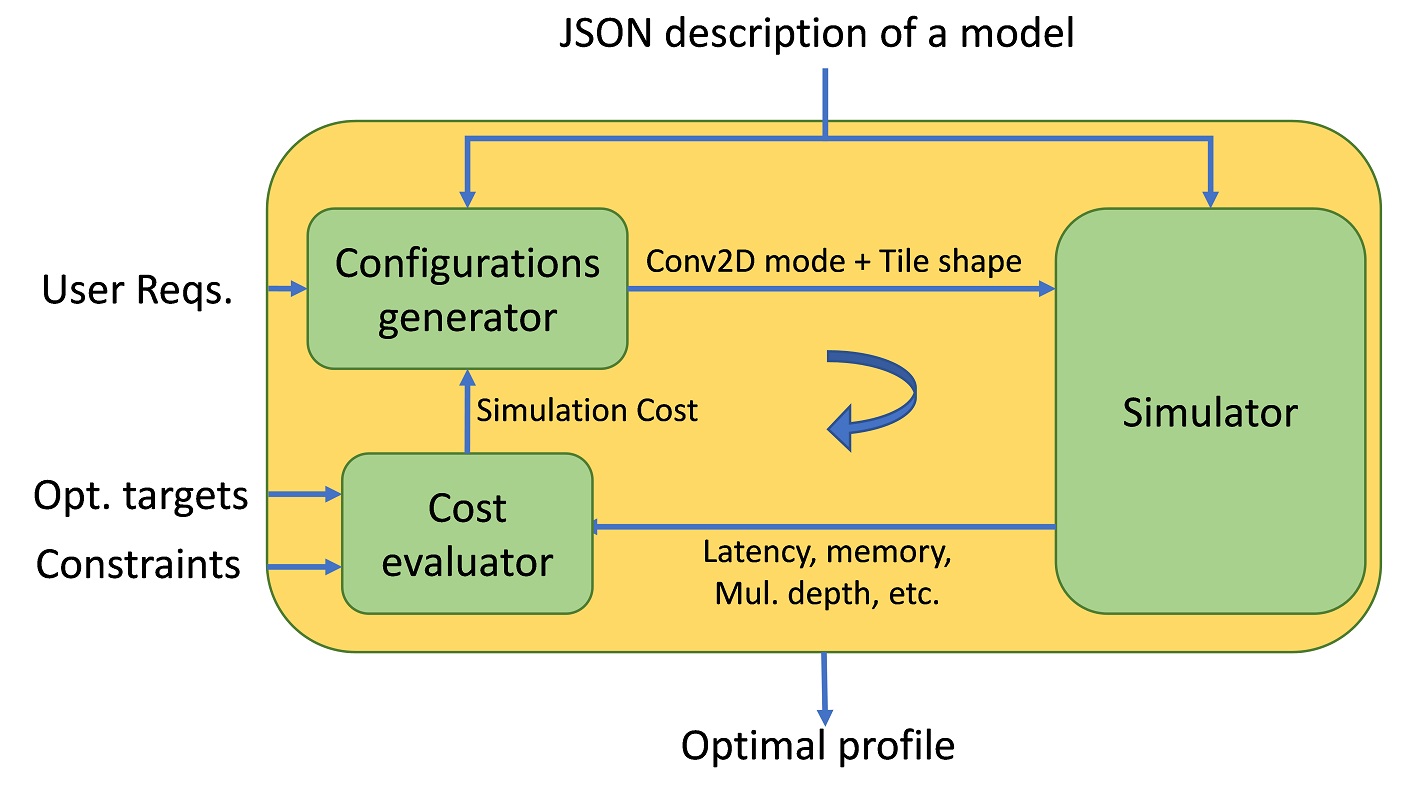}
    \vspace{-21pt}
   \caption{Packing optimizer}
   \vspace{-5pt}
  \label{fig:optimizer}
\end{figure}

Figure \ref{fig:optimizer} presents a schematic of the packing optimizer, and Alg. \ref{alg:optimizer} in Appendix \ref{sec:optalg} presents the process using pseudo code.
It contains three main units: the configuration generator, the cost evaluator, and the simulator.
The user provides a JSON file $m$ that contains the model architecture. 
The configuration generator generates a list of all possible configurations ($\operatorname{GenerateConfs}$ at Step 4, see also Alg \ref{alg:genconf} in Appendix \ref{sec:optalg}), including the packing details and \gls{HE} configuration details  applicable for this architecture.
The simulator unit tests every such configuration and outputs the following data ($res$ at Step 6) for each: the computation time of the different stages including encrypting the model and input samples, running inference, and decrypting the results; the throughput; the memory usage of the encrypted model; input; and output; and more. 
The optimizer passes this data to the cost evaluator for evaluation (Steps 7-16). Finally, it returns the configuration option that yields the optimal cost to the user (among the tested configurations), together with the simulation output profile (Step 17).

{\bf Configuration generator.}
The configuration generator unit receives the model architecture, and generates all applicable configurations for it. For example, if the model has a single convolutional layer it will generate three basic configurations with three possible convolution implementations: the $im2col$ based method, and the two options of our novel method (see Section~\ref{section:convolution}).
If the model has multiple convolutional layers, the $im2col$ based method will not be applicable.
From each of these three basic configurations the generator will create multiple complete configurations by exploring all possible tile shapes. The generator explores possible tile shape using one of two strategies. The first involves brute forcing over all valid options for tile shapes (Alg. \ref{alg:optimizer}). Since these may be numerous, a second strategy searches using a ``steepest ascent hill climbing'' local search algorithm, which requires replacing Step 4 in Alg. \ref{alg:optimizer} with an adaptive $\operatorname{GenerateConf}$ function call.

The local search starts with a balanced tile shape, where the number of slots in every dimension is of the same order. This is a heuristic designed to avoid evaluating tile shapes that are likely to be computationally costly at the beginning of the search. We then iteratively evaluate all the neighbor tile shapes of the current shape and continue to the best-improving neighbor as long as one exists. 
We consider two tile shapes as neighbors if we can obtain one shape from the other by multiplying or dividing the size of some of its dimensions by two. It is possible to use other small multipliers and dividers, but these may involve unnecessary complex computations. We consider one shape as better than another shape based on the costs received from the cost evaluator. Using the local search algorithm highly speeds up the search process and we found empirically that it often results in a global optimum. This was the case in our AlexNet, SqueezeNet-CIFAR and CryptoNets benchmarks.

{\bf Simulator.} The simulator receives as inputs the model architecture and a configuration option. At this stage, we can evaluate the configuration by running it on encrypted input under \gls{HE}. To reduce computational costs, the simulator uses pre-calculated benchmark values such as the CPU time of every HE operation and the memory consumption of a tile (i.e., the memory consumption of a single ciphertext). Then, it evaluates the model on mockup tile tensor objects using these benchmarks. These mockup tile tensors contain only meta data and gather performance statistics. Using this approach, the simulator can simulate an inference operation several order-of-magnitudes faster than when running the complete model on encrypted data. Appendix~\ref{sec:optres} reports the simulator accuracy on AlexNet.
We stress that performing an analytical evaluation instead of an empirical one over a large \gls{NN} requires considering many parameters, which is eventually equivalent to running our simulator using the mockup tile tensors. To allow scientific comparisons of different packings, the simulator also outputs the total number of rotations and multiplications per evaluated circuit  and per layer.

{\bf Cost evaluator.}
The cost evaluation unit evaluates the simulator output data considering the constraints and optimization targets provided by the user. After testing all possible configurations, the highest scoring configuration(s) is sent back as output to the user.

{\bf Evaluating the optimizer performance.}
To demonstrate the advantage of using both the local search algorithm and the simulator, we performed experiments using AlexNet (see Section~\ref{subsec:exp_alexnet} and Appendix \ref{app:exps}). Here, we fixed the number of slots to $16{,}384$, the minimal feasible size for a NN that deep, and set the batch size to 1. The number of configuration options was 1360, with 680 different tile shapes for each convolution packing method. An exhaustive search that uses simulations took 5.1 minutes. In contrast, the local search algorithm took only 6.4 seconds and returned the same result. It did so after evaluating only 40 tile shapes. 

Running the local search method on actual encrypted data took 9.95 hours. Using the simulator time estimations, we predict that exhaustive search on encrypted data would take $\sim 167$ days (assuming unlimited memory). This demonstrates the importance of the mockup-based simulator.

\section{Experimental Results}
\label{section:nn}

Our experiments involve the model weights of a small \gls{NN}  
(CryptoNets \cite{CryptoNets2016}) and a large \gls{NN} (AlexNet \cite{AlexNet2012}) that we trained on the MNIST \cite{lecun1998mnist}, and COVIDx CT-2A \cite{gunraj2021covid} data sets, respectively. We report the results of performing model inference using these weights in encrypted and unencrypted forms. We use AlexNet to demonstrate the power of our framework and CryptoNets to demonstrate the effect of different packing on the computation performance and memory. 
Technical details of the environment we used for the experiments are described in Appendix~\ref{app:exps}. 

\subsection{CryptoNets}
\label{subsection:cryptonets}

The CryptoNets \cite{CryptoNets2016} architecture and the \gls{HE} parameters that we use in our experiments are described in Appendix~\ref{app:nn}. Generally speaking, this network has a convolutional layer followed by two fully connected layers.

Table~\ref{tab_inference1} reports the latency and memory usage for performing a model inference with different tile shapes when $t_3=b=1$. For brevity, we only consider $t_1$ to be at the extreme points (e.g., $t_1=1$, $8{,}192$) or $t_1$ value that led to best performing solution, and some additional samples. The best latency and memory usage are achieved for $t_1=32$, which allows packing the tensors $I,F,W_1$ using the minimal number of tiles. 

Table~\ref{tab_inference2} reports the latency, amortized latency, and memory usage for performing a model inference with different $t_3=b$ values. For every such value, we only report the $t_1,t_2$ values that led to the optimal solutions. Unlike the case where $b=1$, here every choice of $t_3$ leads to a different trade-off between the performance measures.
For example, when increasing $t_3$, the latency and memory consumption increase, but the per-sample amortized latency decreases.
The encryption and decryption time also increase with $t_3$, except for the case $t_3=8{,}192$, where we use the na\"ive SIMD convolution operator.

\subsection{AlexNet Benchmark}
\label{subsec:exp_alexnet}
For this benchmark, we used a variant of AlexNet network~\cite{AlexNet2012} that includes 5 convolution layers, 3 fully connected layers, 7 ReLU activations, 3 BatchNormalization layers, and 3 MaxPooling layers. Following~\cite{towardsalexnet2021, baruch2021fighting}, we created a CKKS-compliant variant of AlexNet by replacing the ReLU and MaxPooling components with a quadratic polynomial activation and AveragePooling correspondingly along with some additional changes.
We trained and tested it on the COVIDx CT-2A dataset,
an open access benchmark of CT images designed by \cite{gunraj2021covid}. We resized the images to $224 \times 224 \times 3$ to fit the input size expected by AlexNet.
See more details in Appendix~\ref{app:nn}.

\begin{table}[t!]
\begin{center}
\caption{
Running a model inference with different tile shapes $[t_1, t_2, t_3]$ when $t_3=b=1$.
The reported values are: the inference latency, the encryption and decryption time, and the memory usage peak.}
\label{tab_inference1}
\begin{tabular}{rrrccc}
\multicolumn{1}{c}{$\mathbf{t_1}$} &
\multicolumn{1}{c}{$\mathbf{t_2}$} &
\multicolumn{1}{c}{$\mathbf{t_3}$} &
\multicolumn{1}{c}{\textbf{Latency}} &
\multicolumn{1}{c}{\textbf{Enc+Dec}} &
\multicolumn{1}{c}{\textbf{Memory}} \\
& & &
\multicolumn{1}{c}{\textbf{(sec)}} &
\multicolumn{1}{c}{\textbf{(sec)}} &
\multicolumn{1}{c}{\textbf{(GB)}} \\
\hline
1 & 8,192 & 1 & 0.86 & 0.04 & 1.58 \\
8 & 1,024 & 1 & 0.56  & 0.04 & 0.76 \\
\textbf{32} & \textbf{256} & \textbf{1} & \textbf{0.56}  & \textbf{0.04} & \textbf{0.73} \\
64 & 128 & 1 & 0.57  & 0.04 & 0.77 \\
128 & 64 & 1 & 0.61  & 0.04 & 0.94 \\
256 & 32 & 1 & 0.68  & 0.05 & 1.37 \\
1,024 & 8 & 1 & 1.93  & 0.14 & 3.17 \\
8,192 & 1 & 1 & 11.10 & 0.80 & 14.81 \\
\end{tabular}
\end{center}
\end{table}

\begin{table}[t!]
\begin{center}
\caption{Running a model inference with different tile shapes $[t_1, t_2, t_3]$, reporting only the optimal $t_1$ and $t_2$ choices for a range of different $t_3=b$ values.
The reported values are: the inference latency, the amortized latency (latency/$b$), the encryption and decryption time, and the memory usage peak.}
\label{tab_inference2}
\setlength{\tabcolsep}{3pt}
\begin{tabular}{rrr....}
\multicolumn{1}{c}{$\mathbf{t_1}$} &
\multicolumn{1}{c}{$\mathbf{t_2}$} &
\multicolumn{1}{c}{$\mathbf{t_3}$} &
\multicolumn{1}{c}{\textbf{Latency}} &
\multicolumn{1}{c}{\textbf{Amortized}} &
\multicolumn{1}{c}{\textbf{Enc+Dec}} &
\multicolumn{1}{c}{\textbf{Memory}} \\
& & &
\multicolumn{1}{c}{\textbf{(sec)}} &
\multicolumn{1}{c}{\textbf{Latency (sec)}} &
\multicolumn{1}{c}{\textbf{(sec)}} &
\multicolumn{1}{c}{\textbf{(GB)}} \\
\hline
32 & 256 & 1 & 0.56 & 0.56 & 0.04 & 0.73 \\
16 & 128 & 4 & 0.56 & 0.14 & 0.05 & 1.20 \\
8 & 64 & 16 & 0.6 & 0.037 & 0.10 & 2.49 \\
4 & 32 & 64 & 0.95 & 0.015 & 0.24 & 6.62 \\
1 & 32 & 256 & 1.94 & 0.008 & 0.70 & 16.38 \\
1 & 8 & 1,024 & 5.6 & 0.0055 & 2.68 & 61.45 \\
1 & 2 & 4,096 & 21.57 & 0.0053 & 12.55 & 242.46 \\
1 & 1 & 8,192 & 41.32 & 0.005 & 1.29 & 354.47 \\
\end{tabular}
\end{center}
\vspace{-15pt}
\end{table}

\begin{table}[t!]
\setlength{\tabcolsep}{2pt}
\begin{center}
\caption{AlexNet executed in our framework with different configurations. See configuration description in Section \ref{subsec:exp_alexnet} }
\label{tab_alex_he}
\begin{tabular}{lcccc}
\multicolumn{1}{c}{\textbf{Config.}}
&
\multicolumn{1}{c}{\textbf{Latency}} &
\multicolumn{1}{c}{\textbf{Amortized}} &
\multicolumn{1}{c}{\textbf{Enc+Dec}} &
\multicolumn{1}{c}{\textbf{Memory}}  \\
&
\multicolumn{1}{c}{\textbf{(sec)}} &
\multicolumn{1}{c}{\textbf{Latency (sec)}} &
\multicolumn{1}{c}{\textbf{(sec)}} &
\multicolumn{1}{c}{\textbf{(GB)}}  \\

\hline
PT-Latency & 298.8 & 298.8 & 5.289 & 163 \\
PT-TP      & 785   & 98.1  & 5.102 & 604 \\
CT-Latency & 350.4 & 350.4 & 5.018 & 252 \\
CT-TP      & 804.8 & 201.2 & 5.671 & 739 \\
\end{tabular}
\end{center}
\end{table}

For the convolutional layers, we used the packing methods described in Section~\ref{subsec:conv_sequence}. The biases were packed in $5$-dimensional tile tensors with compatible shapes, allowing us to add them to the convolution outputs. 
The fully connected layers were handled using the matrix-matrix multiplication technique of Section~\ref{subsection:high_level_ops}. The input to these layers arrives from the convolutional layers as a 5-dimensional tile tensor, $[\frac{*}{t_1}, \frac{1\sim}{t_2}, \frac{1\sim}{t_3}, \frac{9216}{t_4}, \frac{b}{t_5}]$. Therefore, the first fully connected layer is packed in $5$ dimensions as well: $[\frac{4{,}096}{t_1}, \frac{1\sim}{t_2}, \frac{1\sim}{t_3}, \frac{9216}{t_4}, \frac{*}{t_5}]$. Its output, $[\frac{4{,}096}{t_1}, \frac{1\sim}{t_2}, \frac{1\sim}{t_3}, \frac{1?}{t_4}, \frac{b}{t_5}]$, is replicated along dimensions $2$ through $4$, then flattened using the flatten operator to\\ $[\frac{4{,}096}{t_1}, \frac{*}{t_2 t_3 t_4},  \frac{b}{t_5}]$, from which we can continue normally.

We measured the accuracy of running vanilla AlexNet
\cite{AlexNet2012} and the HE-friendly AlexNet \cite{baruch2021fighting} using PyTorch\footnote{PyTorch library 1.5.1 \url{https://pytorch.org}} over a plaintext test-set. The results were $0.861$ and $0.806$, respectively. We did not observe additional accuracy degradation when running the HE-friendly AlexNet using our framework over encrypted data. We emphasize that the above accuracy-drop results from using HE-friendly NN and not from using our framework. We expect that future AI improvements will close this gap by offering improved HE-friendly NNs.

Table~\ref{tab_alex_he} reports the time and memory consumption for the latter experiment using $4$ configurations on a set of $10b$ representative samples. The configurations involve unencrypted model weights (\textit{PT}) and encrypted model weights (\textit{CT}) optimized for low latency (\textit{Latency}) or high throughput (\textit{TP}). For these configurations, we also compared the inference results with the inference results of running HE-Friendly AlexNet on PyTorch over the plaintext test-set by calculating the Root Mean Square Error (RMSE). These were always less than $4\mathrm{e}{-3}$.

\section{Comparison with State-of-the-Art}
\label{section:comparison}

\subsection{Matrix Multiplication}
\label{sec:matrix_compare}

Table \ref{table:compare-matrix} compares our tiles-tensor-based matrix-multiplication technique for multiplying two $d\times d$ matrices with the state-of-the-art. We keep the matrices in tile tensors of shapes $[\frac{d}{\sqrt{d}}, \frac{d}{d}, \frac{*}{\sqrt{d}}]$ and $[\frac{*}{\sqrt{d}}, \frac{d}{d}, \frac{d}{\sqrt{d}}]$. Similar to \cite{faster_linear_transformations}, we assume the na\"ive technique uses the na\"ive $\BigO(d^3)$ algorithm over ciphertexts that hold one value each. 
Some previous works bound the dimension $d$ by some function of $s$. In \cite{faster_linear_transformations}, the authors assume $d<s$, and in \cite{secure_outsourced}, the authors assume that $d<\sqrt{s}$. The latter bound was also used in \cite{secure_outsourced_malicious_mpc} for a version addressing a malicious adversary, and in \cite{secure_outsourced_multikey} for a version addressing multi-key encryption. In both cases, $s$ can be made arbitrarily large in two ways. We explain and compare these works to our technique, which does not place any bound on $d$.

{\bf Increasing the scheme's parameters.} By increasing the parameters of the scheme, $s$ can be made arbitrarily large. However, this increases the relative overhead of every \gls{HE} operation, i.e., increasing $s$ by a factor  $\alpha$ increases the time of every \gls{HE} operation by a factor $\alpha'>\alpha$. Moreover, in practice, there is a maximal setting that every \gls{HE} library supports. Our technique is superior because it enables us to use the smallest $s$ allowed by the security requirement for the multiplication.

{\bf Arbitrarily simulating  many slots.} An array $c$ of $d^2 > s$ can be simulated na\"ively using $t = \lceil \frac{d^2}{s} \rceil$ ciphertexts. In this simulation, each operation performed on $c$ (especially rotations) is translated to $\BigO(t)$ operations on ciphertexts. This makes the complexity deteriorate to $\BigO(\frac{d^3}{s})$. With our technique, we set the shape of the matrices to $[\frac{d}{s^{1/4}}, \frac{d}{s^{2/4}}, \frac{*}{s^{1/4}}]$ and $[\frac{*}{s^{1/4}}, \frac{d}{s^{2/4}}, \frac{d}{s^{1/4}}]$. This yields a complexity of $\BigO(\frac{d^3}{s})$, which is similar to that of \cite{secure_outsourced}.But again, our technique has a lower multiplication depth because it does not involve masking.

When considering complete \gls{NN}, the authors of \cite{secure_outsourced} argue that it is possible to convert fully-connected and convolutional layers to matrix-matrix multiplications. In that sense, our method, as shown above, are better.

\begin{table}[t!]
\caption{A comparison of techniques to multiply two $d\times d$ matrices using ciphertexts with $s$ slots. The columns represent the multiplication complexity, and the multiplicative-depth of each technique.
We assume the shape of the matrices in our technique are $[\frac{d}{\sqrt{d}}, \frac{d}{d}, \frac{*}{\sqrt{d}}]$ and $[\frac{*}{\sqrt{d}}, \frac{d}{d}, \frac{d}{\sqrt{d}}]$.
}
\label{table:compare-matrix}
\begin{tabular}{llcc}
Method & Complexity & Mul-Depth \\
\hline
Na\"ive                                                                      & $\BigO(d^3)$   & 1                         \\

\cite{faster_linear_transformations, faster_linear_transformations_revisited}$^*$                & $\BigO(d^2\ceil{\frac{d}{s}})$   & 1                         \\
\cite{secure_outsourced}$^*$       & $\BigO(d\ceil{\frac{d^2}{s}})$     & 3                         \\
Ours                          & $\BigO(d\ceil{\frac{d^2}{s}})$     & 1 \\
\hline
\multicolumn{3}{l}{%
        \begin{minipage}{6.5cm}%
            \footnotesize{$^*$ \cite{faster_linear_transformations,faster_linear_transformations_revisited} assumed $d<s$ and \cite{secure_outsourced} assumed $d^2<s$. In both cases, supporting arbitrarily large $d$ can be done by increasing $s$ or by simulating a large slot number. The first option increases the time to compute every HE operation. Moreover, current implementations of HE-library support only limited values of $s$. The latter option replaces each operation by $\BigO(t)$ operations on ciphertext, where $t=\lceil\frac{d}{s}\rceil$ for \cite{faster_linear_transformations, faster_linear_transformations_revisited} and $t=\lceil\frac{d^2}{s}\rceil$} for \cite{secure_outsourced}.
            
        \end{minipage}%
    }\\
\end{tabular}
\vspace{-10pt}
\end{table}

\subsection{Convolution}
\label{subsection:related_work_conv}

The state-of-the-art in implementing convolutional layers includes \cite{GAZELLE2018, GALA, HEAR}. We compare these  to our implementation by distinguishing  between two cases: simple single-input-single-output (SISO) and multiple-input-multiple-output (MIMO). 

\subsubsection{SISO}

The simple SISO case involves one image, one input channel, and one filter, $b=c=f=1$. Here, all previous approaches pack the image into a single ciphertext and compute the convolution using $w_F h_F - 1$ rotations (excluding the 0 rotation) and $w_F h_F$ multiplications. However, the assumption that the image fits in a single ciphertext limits the maximal allowed image size. As mentioned above, all \gls{HE} libraries limit the size of their ciphertexts. For example, SEAL \cite{sealcrypto} limits the ciphertext slots to 16,384, which is not large enough to contain $224\times 224$ images such as those required for AlexNet. While it is possible to simulate a large ciphertext (see Section \ref{sec:matrix_compare}), this increases the multiplication depth by at least  $1$ per convolutional layer. For deep networks such as AlexNet, which have $5$ convolutional layers, the extra cost of $5$ multiplication levels makes the computation impossible when using libraries such as SEAL, where the maximal depth is bounded and bootstrapping is not supported. Our approach captures the method above as a special case where  $w_I \leq t_1$ and $h_I\leq t_2$, i.e., the image fits inside a single tile. Our generalization offers two advantages.

First, it allows the network to efficiently handle images larger than the maximal ciphertext size; this was not described in previous works. As mentioned above, na\"ively simulating larger ciphertexts using $n$ smaller ones requires $\BigO(n)$ rotations and increases the multiplication depth. In contrast,  
in our method, if we divide the image into $t$ square tiles, it multiplies the number of rotations by $\BigO(\sqrt{n})$ instead of $\BigO(n)$ and no extra multiplication levels are consumed. Lemma~\ref{lem:conv1} shows that the number of rotations is $\BigO(w_F \ceil{\frac{w_I}{t_1}} + h_F \ceil{\frac{h_I}{t_2}} + w_F h_F) \text{ rotations}$.

Second, our method makes it beneficial to split even medium size images into smaller tiles. This improves performance on three accounts: 1) smaller ciphertexts are more efficient per-slot, i.e., if an operation on a ciphertext of $s$ slots costs $t$ time, then the same operation on a ciphertext of $2s$ slots costs more than $2t$ time; 2) as mentioned above, the number of rotations increases sub-linearly; 3) many small operations are easier to parallelize than a few large operations.

\begin{table}[ht!]
\caption{Comparing time and number of operations for a SISO ($b=c=f=1$) convolution benchmark of image of size $128\times 128$, filter size $3\times 3$, and different tile sizes. Each row has a different setting for $t_1,t_2$, and  $t_3=t_4=t_5=1$ (see Section~\ref{section:convolution}). The first row captures as a special case existing SISO methods~\cite{GAZELLE2018, GALA, HEAR}. Subsequent rows demonstrate the performance advantage of our generalized method. 
}
\label{table:compare-conv}
\begin{tabular}{cccccc}
$t_1$ & $t_2$& Tiles & Time (sec) & Rotations & Multiplications  \\
\hline
128 & 128 & 1 & 0.119 & 8 & 9 \\
64 & 128 & 2 & 0.064 & 10 & 18\\
64 & 64 & 4 & 0.030 & 12 & 36\\
32 & 64 & 8 & 0.019 & 16 & 72
\end{tabular}
\end{table}

Table~\ref{table:compare-conv} demonstrates this advantage. It shows a SISO convolution computed with different ciphertext sizes, starting with $16{,}384$ slots shaped as a $128\times 128$ tile. The computations go down to the smallest ciphertext of $2{,}048$ slots,  which still supports secure computation. The table shows the speedup (more than $6\times$), the sub-linear growth in the number of rotations, and the linear growth in the number of multiplications. If more computation depth is required, the ciphertexts cannot be made smaller due to security constraints. In this case, we can reduce $t_1$ and $t_2$, while increasing $t_5$, the batch dimension. Table~\ref{table:compare-conv} will stay the same, except the rows will support batch sizes of $1,2,4,8$ respectively, with amortized time per sample of $0.119, 0.068, 0.061, 0.016$ seconds.

\subsubsection{MIMO}

In MIMO, the number of channels $c$ and filters $f$ is larger than $1$.

As a special case, the tile tensor and its accompanying notation captures the packing options described in \cite{GAZELLE2018, HEAR, GALA}, which is $[\frac{w_I}{t_1},\frac{h_I}{t_2},\frac{c}{t_3}]$. Previous approaches restricted $w_I\leq t_1$, $h_I\leq t_2$, but $c$ and $t_3$ can be set arbitrarily. If $c\leq t_3$ then all channels are packed into a single ciphertext. If $t_3=1$, then each channel is in a separate ciphertext.

In HEAR~\cite{HEAR} an additional packing optimization is described. They note that after a mean pooling layer with strides, the output comes out with strides $\delta$, i.e., only the pixels at $x$ and $y$ coordinates such that $x\bmod \delta=0$ and $y\bmod \delta=0$ are populated with actual pixels. They use the vacant pixels to pack more channels. Using tile tensor shapes this is equivalent to $\left[\frac{w_I}{w_I},\frac{\delta}{\delta},\frac{h_I}{h_I},\frac{\delta}{\delta},\frac{c}{t_3}\right]$, where $w_I,h_I$ are the dimensions of the output image, and the vacant slots are represented by two new dimensions, such that $w_I\delta=t_1$ and $h_I\delta=t_2$. We fill them with channel data, having a total capacity of $\delta^2 c$ channels. Packing those channels together and then summing over them requires processing time. We note that the interleaved packing method described in Section~\ref{section:convolution} reduces the need for such a packing optimization. As with interleaved packing, if $\delta$ divides the number of tiles along a dimension, the stride can be implemented by simply discarding tiles; this saves computation time and does not create vacant pixels.

Previous approaches described numerous techniques for exploiting hoisted rotations, and for optimizing the number of rotations in MIMO convolutions. Since all existing packing methods are a special case of tile tensor capabilities, these methods are applicable within our framework. Integrating them within our optimizer is reserved for future work. 

\subsection{Neural Network Inference}
\label{subsection:related_work_nn}

\begin{table*}[t!]
    \caption{Scheme comparison. (NI) - Non interactive. (PL) - reported on practical latency of less than 2 hours. (LM) - Reported support in latency vs memory tradeoff. (ALL) - Reported support in amortized-latency vs. latency tradeoff. (EM) - Reported data for encrypted model. (SB) - Security bits. Larger image size involves more complex computations.}
    \label{tab:schemes}
    \begin{center}
    \begin{tabular}{lclclcclr}
    \hline
    Scheme Name & 
    NI &
    PL &
    LM &
    ALL &
    EM &
    SB &
    Evaluated Networks &
    Image size\\
    \tablespace
    \hline
    \tablespace
    GAZELLE \cite{GAZELLE2018} & \priority{0} & \priority{100} & \priority{0} & \priority{0} & \priority{0}& \textbf{128} & CryptoNets & $32 \times 32 \times 3$  \\
    
    Jiang et al. \cite{secure_outsourced} & \priority{100} & \priority{100} & \priority{0} & \priority{0} & \priority{0}& \textbf{128} & CryptoNets & $28 \times 28 \times 1$  \\

    nGraph-HE2 \cite{HET}  & \priority{0} &  \priority{100} & \priority{0} & \priority{0} & \priority{0} & \textbf{128} & CryptoNets, MobileNetV2 & $\mathbf{224 \times 224 \times 3}$  \\
    
    Lloret-Talavera et al. \cite{optane}  & \priority{0} &  \priority{50}$^*$ & \priority{0} & \priority{0} & \priority{0} & \textbf{128} & MobileNetV2, ResNet-50 & $\mathbf{224 \times 224 \times 3}$  \\
    
    SeaLION \cite{sealion} & \priority{0} & \priority{100} & \priority{0} & \priority{0} & \priority{0} & \textbf{128} &  DNN-30, DNN-100, CNN-16 & $28 \times 28 \times 1$  \\
    
    TenSEAL \cite{tenseal} & \priority{100} & \priority{100} & \priority{0} & \priority{0} & \priority{0}& \textbf{128} & CryptoNets & $28 \times 28 \times 1$  \\

    DOReN \cite{Doren} & \priority{100} & \priority{0} & \priority{0} & \priority{0} & \priority{0} & \textbf{128} & VGG7, ResNet20, Lola, GHE, SHE & $32 \times 32 \times 3$ \\

    REDsec \cite{REDsec} & \priority{100} & \priority{0} & \priority{0} & \priority{0} & \priority{0} & 80 & BAlexNet$^{**}$ & $\mathbf{224 \times 224 \times 3}$   \\
    
    Lee et al. \cite{ppml22} & \priority{100} & \priority{0} & \priority{0} & \priority{0} & \priority{0} & 111.6 & ResNet-20 & $32 \times 32 \times 3$   \\
    
    CHET \cite{chet_compiler} & \priority{100} & \priority{100} & \priority{100} & \priority{50}$^\bot$ & \priority{0} & \textbf{128} & LeNet-5, CIFAR-Squeezenet, Industrial & $32 \times 32 \times 3$  \\
    
    HEMET \cite{hemet} & \priority{100} & \priority{100} & \priority{0} & \priority{0} & \priority{0} & \textbf{128} & AlexNet, CIFAR-Squeezenet, InceptionNet & $32 \times 32 \times 3$  \\

    \textbf{HeLayers (ours)} & \priority{100} & \priority{100} & \priority{100} & \priority{100} & \priority{100} & \textbf{128} &  AlexNet, CIFAR-Squeezenet, CryptoNets & $\mathbf{224 \times 224 \times 3}$ \\
    \multicolumn{9}{l}{%
        \begin{minipage}{12cm}%
            \footnotesize{$^*$ The reported latency is 63 hours for a batch of 2048 images (1.84 minutes of amortized latency).}
    
            \footnotesize{$^{**}$ BinaryAlexNet with binary weights and accuracy drop to 61.5\%}
            
            \footnotesize{$^{\bot}$ CHET supports batching but the paper does not provide a description, analysis, or measurements.}
            
        \end{minipage}%
    }\\
    \end{tabular}
    \end{center}
\end{table*}

Table \ref{tab:schemes} compares our framework, experiments, and results with prior art. The first column (NI) indicates whether the framework supports non-interactive secure computations. This is a major advantage that HeLayers provides over GAZELLE \cite{GAZELLE2018}, nGraph-HE \cite{HET}, SeaLion \cite{sealion}, etc. In general, using the client to assist the server reduces the latency, which in turn enables the evaluation of deep networks such as ResNet and VGG. However, it misses the paradigm of delegating the entire computations to the cloud. 

The second column (PL) indicates whether the reported experiments showed practical inference latency of less than two hours. We use this Boolean comparison metric instead of comparing exact latencies because different frameworks reported their measurements on different network architectures, over different datasets with images of different sizes when using different platform configurations such as number of CPUs and memory size. Furthermore, some of the schemes e.g., CHET \cite{chet_compiler} and SeaLION \cite{sealion} are not freely available online nor do they provide open source code that we can evaluate, where implementing their code from scratch may result in a biased implementation. Using the PL metric allows us to distinguish HeLayers from some works, while we use other comparison metrics to show HeLayers' superiority over the other solutions.

The following two columns indicate whether the papers reported on a mechanism or a method to evaluate and select the packing choice that mostly fits the given criteria. Specifically, it indicates whether they support tradeoffs between latency versus memory (LM) and amortized-latency versus latency (ALL). While most schemes considered only one type of packing, as far as we know only CHET suggested using different packing and selecting among them according to some criterion such as latency, energy, or memory. However, they did not present any amortized latency tradeoffs in their paper and they did not discuss how this selection mechanism works.

The other comparison metrics we considered include whether the model is encrypted before the inference evaluation, whether the security strength was at least 128 bits, and whether the image size was at least 224x224x3. All of these parameters highly affect the performance of the experimental evaluation. It can be observed from the table that HeLayers is the only framework that provides a non-interactive solution over large images and an encrypted model that considered different trade-offs and reported practical results that measure less than several minutes.

When considering the different packing proposals. TenSEAL \cite{tenseal} uses diagonalization techniques for matrix-multiplication and im2col for convolution, assuming a single, image as input. nGraph-HE2 \cite{HET} uses \gls{SIMD} packing, which is a special case of our framework when optimized for the largest possible batch size. The closest work to ours is CHET \cite{chet_compiler}, which uses a similar approach of an abstract data structure, CipherTensor, combined with automatic optimizations. We believe CipherTensors are less flexible than tile tensors. They include a  
small fixed set of implemented layouts, each with its kernel of algorithms, whereas tile tensors offer a wider variety of options with a single set of generalized algorithms. Further, it was not demonstrated that CipherTensors offer an easy method to trade latency for throughput and control memory consumption, as is possible in tile tensors by controlling the batch dimension. Finally, CipherTensors require replication of the input data using rotations, whereas some of these replications can be avoided using tile tensors. It might be the case that by this time CHET has been enhanced to include other optimizations, but we only compare our solution to data that is publicly available, i.e., to \cite{chet_compiler}.

\begin{table}[t!]
    \caption{Comparison of the SqueezeNet-CIFAR benchmark with our tile tensor framework and CHET \cite{chet_compiler}.}
    \label{tab:squeezenetcompare}
    \centering
    \begin{tabular}{lcc}
    \textbf{Framework} & \textbf{Latency} & \textbf{Amortized} \\
    &  \textbf{(sec)} & \textbf{Latency (sec)} \\
    \hline
    CHET \cite{chet_compiler} ($b=1$) & 164.7 & 164.7 \\
    Ours-PT ($b=1$) & 131.8 & 131.8 \\
    Ours-CT ($b=1$) & 143.5 & 143.5 \\
    \hline
    Ours-PT ($b=64$) & 913.1 & 14.26 \\
    Ours-CT ($b=64$) & 1214.2 & 18.971 \\
    \end{tabular}
\end{table}

CHET is the closest framework to ours, and even though it does not provide a freely available solution or an open source one, we attempted below to provide as fair a comparison as possible to the latency reported in \cite{chet_compiler}. The authors of CHET demonstrated its efficiency for large networks by using SqueezeNet-CIFAR, which is a reduced \cite{squeeze0} SqueezeNet \cite{squeezenet} model over CIFAR-10.
It has at least $50 \times$ fewer parameters than AlexNet but it is a bit deeper (23 layers instead of 17). see Appendix \ref{app:nn}. We evaluated SqueezeNet-CIFAR on HeLayers and report the results in Table \ref{tab:squeezenetcompare}. For a fair comparison with CHET, we reduced the number of cores to 16 while disabling hyper-threading as in CHET. In addition, the SqueezeNet-CIFAR repository \cite{squeeze0} contains two models, we used the larger model among the two (the smaller model has $1.2\times$ less latency than the larger model).
When optimizing for latency, our framework is $1.24 \times$ faster than CHET. In addition, we also provide the results when optimizing for throughput and for running with encrypted models, which show even better amortized results (an $11.5 \times$ speedup over CHET).

\section{Conclusions}
\label{section:conclusions}

We presented a framework that acts as middleware between \gls{HE} schemes and the high-level tensor manipulation required in AI.

Central to this framework is the concept of the tile tensor, which can pack tensors in a multitude of ways. The operators it supports allow users to feel like they are handling ordinary tensors directly. Moreover,  the operators are implemented with generic algorithms that can work with any packing arrangement chosen internally.

The optimizer complements this versatile data structure by finding the best configuration for it given the user requirements and preferences. We demonstrated how this approach can be used to improve latency for small networks, adapt to various batch sizes, and scale up to much larger networks such as AlexNet.

Our tile tensor shape notation proved very useful for both research and development. 
Having the notation used in debug prints and error messages, configured manually in unit tests, and printed out in the optimizer log files, helped reduce development cycles considerably. Also, in this paper we used it to concisely and accurately describe  complicated computations (e.g., Figure~\ref{fig:cryptonets}).
We hope the community will adopt it as a standard language for describing packing schemes.

Our framework is the first to report successful and practical inference over a large (in terms of HE) NN such as AlexNet over large images. Today, we are in the process of extending our library to support bootstrap capabilities. Combining this and our framework should enable us to run even larger networks such as VGG-16 and GoogleNet, which until now were only reported for client-aided designs or for non-practical demonstrations.

\begin{acks}
This research was conducted while the authors worked at IBM Research - Israel. Other than that, this research received no specific grant from any funding agency in the public, commercial, or not-for-profit sectors.
\end{acks}

\bibliography{main}
\bibliographystyle{ACM-Reference-Format}

\appendix

\section{Experiments setup}
\label{app:exps}

For the experiments we used an
Intel(R) Xeon(R) CPU E5-2699 v4 @ 2.20GHz machine with 44 cores (88 threads) and 750GB memory. Unless specified otherwise, we used only 40 threads and avoided hyper-threading by instructing the OpenMP library to pin one software thread per core. We use the CKKS SEAL~\cite{sealcrypto} implementation targeting $128$ bits security, and all the reported results are the average of $10$ runs. 

\subsection{Neural Networks}\label{app:nn}

\begin{table*}[ht!]
    \centering
    \caption{Network architecture used in our experiments. C, A, P, D, and F denote convolution, activation, pooling, dense, and fire module layers, respectively. A fire module has the form C-A-C2-A, where C2 represents two concated convolutions.}
    \label{tab:nets}
    \begin{tabular}{l|c}
         Name &  Architecture \\
         \hline
         CryptoNets \cite{CryptoNets2016} & C-D-D \\
         SqueezeNet-CIFAR \cite{squeeze0} (HE-friendly \cite{baruch2021fighting}) &
         C-A-P-F-F-P-F-F-C-A-P\\
         AlexNet \cite{AlexNet2012} (HE-friendly \cite{baruch2021fighting}) & C-A-P-C-A-P-C-A-C-A-C-A-P-D-A-D-A-D \\
         \hline
    \end{tabular}
\end{table*}

Table \ref{tab:nets} summarizes the network architectures that we evaluated in this paper. It includes CryptoNets \cite{CryptoNets2016}, over images of size $28 \times 28$, padded to $29 \times 29$, An HE-Friendly variants \cite{baruch2021fighting} of AlexNet~\cite{AlexNet2012} and SqueezeNet-CIFAR \cite{squeeze0}, which we train using the methods of \cite{baruch2021fighting}.

\begin{figure}[ht!]
\centering
    \centering
    \includegraphics[width=0.99\linewidth]{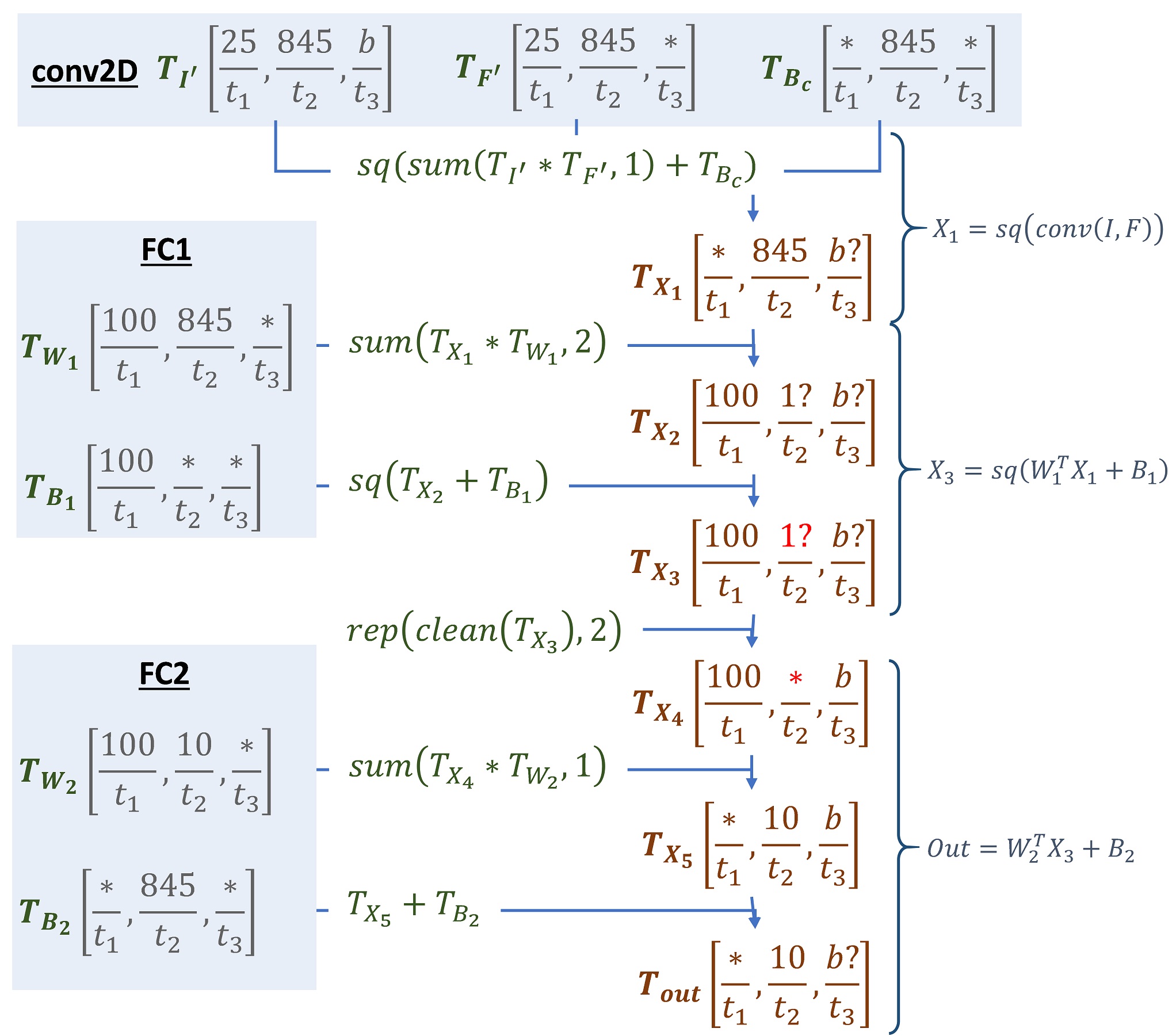}
    \caption{An illustration of our CryptoNets implementation using tile tensors. For simplicity, we only indicate the component-wise square activation layers by the $sq()$ function because they maintain the tile tensor shape. The equations on the right represent the underlying tensor operations. The input tensors are $I, F, B_c, W_1, B_1, W_2, B_2$, where $I',F'=im2col'(I,F)$.} 
    \label{fig:cryptonets}
\end{figure}

\subsubsection{CryptoNets}
\label{appendix:cryptonets}

We implemented the network using tile tensors of shape $\left[\frac{n_1}{t_1}, \frac{n_2}{t_2}, \frac{b}{t_3}\right]$, where $b$ is the batch size. In practice, we only report the results for the case $t_3=b$ that minimizes the overall latency by filling all the ciphertext slots ($8{,}192$ in this case). For the convolution layer, we use the na\"ive SIMD method (Section~\ref{sec:naive_conv}) when $b$ equals the number of plaintext slots and $t_1=t_2=1$. Otherwise, we use our variant of the im2col operator (Section~\ref{subsec:background_im2col}).
These methods work better than our novel convolution operator when the images are relatively small (e.g., MNIST images) and the network has one convolutional layer.

Figure \ref{fig:cryptonets} shows the tile tensor flow in our implementation. Here, the inputs $I$ and $F$ are the image and filter matrices, respectively, and $I',F'=im2col'(I,F)$. In addition, $B_c$ is the trained bias of the convolution layer and $W_1,W_2,B_1,B_2$ are the trained weights and bias info of the \gls{FC} layers.

{\bf CryptoNets HE configurations.}
We set the plaintext poly-degree to $16{,}384$, and set the modulus chain $\{45,35,35,35,35,35,45\}$ when either $t_1=1$ or $t_2=1$. Otherwise, we set the modulus chain to $\{45,35,35,35,35,35,35,45\}$, for increasing the multiplication depth by one, needed for the $clean$ operator (see Section~\ref{subsection:high_level_ops}). 

\subsubsection{AlexNet}
\label{appendix:alexnet}

{\bf Dataset.} The COVIDx CT-2A Data-set
is an open access benchmark of CT images dataset designed by \cite{gunraj2021covid}, that contains three classes of chest CT images: \textit{Normal}, \textit{Pneumonia} or \textit{COVID-19} cases.
For the experiment we took a subset of 10K images per class for training, 1K images per class for validation, and 201 images in total for test with 67 random samples from each class. We chose a small subset to speed up accuracy measurements. 
The images were resized to $224 \times 224 \times 3$ to fit the input size expected by AlexNet.

{\bf Preparing a model for inference over encrypted data.}
The batch normalization that we used for training requires division, which is not a CKKS primitive. Therefore, for the model inference we used a technique similar to~\cite{hebatchnorm2018} to ``absorb'' batch normalization layers into neighboring layers. This was done by modifying the neighbor layer's parameters in such a way that the resulting transformation of the layer is equivalent to a sequential application of batch normalization and the original layer. The resulting network is computationally equivalent, but does not include batch normalization layers. 

{\bf HE configurations}
We set the plaintext poly-degree to $32{,}768$, and use the modulus chain $\{49,39, 39, \ldots, 39,49\}$ of size $22$.

\subsubsection{SqueezeNet-CIFAR}
For SqueezeNet-CIFAR \cite{squeeze0} we used the same techniques that we used for AlexNet. Here, we set the plaintext poly-degree to $32{,}768$, and use the modulus chain $\{50, 30, \ldots, 30, 50\}$ of size $28$.

\section{Tile Tensors Definition}
\label{appendix:tt_def}

\begin{definition}[External tensor]
A $k$-dimensional external tensor $E$ is a $k$-dimensional tensor that each of its elements is itself a $k$-dimensional tensor, all having an identical shape. These internal tensors are referred to as {\em tiles}, their shape as the {\em tile shape}, and the shape of the external tensor as the {\em external shape}. A slot in $E$ is identified by $E(a_1,\ldots,a_k)(b_1,\ldots,b_k)$ where $a_i$ are the external indices of a tile, and $b_i$ are the internal indices inside the tile.
\end{definition}

\begin{definition}[Tile tensor shape]
A $k$-dimensional tile tensor shape is comprised of an external shape $[e_1,\ldots,e_k]$,
tile shape $[t_1,\ldots,t_k]$,
original shape $[n_1,\ldots,n_k]$,
replication counts $[r_1,\ldots,r_k]$,
interleaved Boolean indicator $[l_1,\ldots,l_k]$,
and unknown Boolean indicators $[u_1,\ldots,u_k]$.
It is required that $\forall_i (r_i=1\lor n_i=1) \land (max(r_i,n_i)\leq e_i t_i)$.
\end{definition}

\begin{definition}[External tensor logical indices]
Given a tile tensor shape $S$, and an external tensor $E$, and a specific slot in $E$ specified by external indices $(a_1,\ldots,a_k)$, and internal indices $(b_1,\ldots,b_k)$, then this slot is associated with the logical indices\\ $(c_1,\ldots,c_k)$ with respect to $S$, computed as follows: For $i=1,\ldots,k$, if the interleaved indicator $l_i$ is true, then $c_i=b_i e_i + a_i$ else $c_i=a_i t_i+b_i$.
\end{definition}

\begin{definition}[Validity relation, Packed tensor]
A tile tensor shape $S$ is valid for an external tensor $E$ if their external shapes and tile shapes match, and
there exists a tensor $T[n_1,\ldots,n_k]$ such that for $T_1=broadcast(T,[n_1 r_1,\ldots,n_2 r_2])$ it holds that\\ $E(a_1,\ldots,a_k)(b_1,\ldots,b_k)=T_1(c_1,\ldots,c_k)$ for all slots with internal, external, and logical indices $a_i,b_i,c_i$, such that  $\forall_i c_i\leq n_i r_i$.
For all other slots of $E$, if $\forall_i ((c_i\geq r_i n_i) \rightarrow \lnot u_i))$ then these slots are set to zero.
$T$ is the \emph{packed tensor}.
\end{definition}

\begin{definition}[Tile tensor]
Tile tensor is a pair $(E,S)$ where $E$ is an external tensor and $S$ a tile tensor shape that is valid for it.
\end{definition}

\begin{definition}[Unpack operator]
Given a tile tensor $T_A=(E,S)$ the operator $unpack(E)$ results with the packed tensor of $T_A$.

\end{definition}

\begin{definition}[Pack operator]
Given a tensor $A$ and a tile tensor shape $S$ whose original shape matches the shape of $A$, then the pack operator $pack(A,S)$ results with a tile tensor $T_A=(E,S)$ such that $A$ is the packed tensor of $T_A$.
\end{definition}

\subsection{Tile Tensor Shape Notation}

A tile tensor shape can be specified with a special notation involving a list of symbols. Each element in the list specifies the details of one dimension. $\frac{n_i}{t_i}$ specifies the original and tile shape along this dimension, and $r_i=1, e_i=\ceil{\frac{n_i}{t_i}}, l_i=u_i=false$. $\frac{*r_i}{t_i}$ further specifies the replication count and $n_i=1$, and $\frac{*}{t_i}$ specifies $n_i=1, r_i=t_i$.
$\frac{n_i\sim}{t_i}$ specifies $l_i=true$, and $\frac{n_i\sim e_i}{t_i}$ specifies a value for $e_i$ other than the default mentioned above. For any of the above mentioned options a "?" symbol above the line indicates $u_i=true$.

\subsection{Operators}
\label{subsec:ops_tt_elementwise}

\begin{definition}[Tile tensor shape compatibility]
Tile tensor shapes $S$ and $S'$ are compatible for all $i$, $t_i=t_i'$,
$(n_i=n_i' \land e_i=e_i' \land l_i=l_i') \lor ( n_i=1 \land r_i=t_i)
\lor ( n_i'=1 \land r_i'=t_i')
$.
\end{definition}

\begin{definition}[Tile tensor addition]
Let $T=(E,S)$ and $T'=(E',S')$ be tile tensors with compatible shapes, then $T+T'=T''=(E'',S'')$ such that $E''=E'+E'$, 
$n_i''=max(n_i,n_i')$, $r_i''=min(r_i,r_i')$,
$u_i''=(e_i t_i - n_i r_i \ne e_i' t_i'-n_i'r_i') \lor u_i \lor u_i''$, $l_i''=l_i \lor l_i'$.
\end{definition}

\begin{definition}[Tile tensor elementwise multiplication]
Let $T=(E,S)$ and $T'=(E',S')$ be tile tensors with compatible shapes, then $T*T'=T''=(E'',S'')$ such that $E''=E'*E'$, 
$n_i''=max(n_i,n_i')$, $r_i''=min(r_i,r_i')$,
$u_i''=((e_i t_i - n_i r_i = e_i' t_i'-n_i'r_i') \land u_i \land u_i') \lor ((e_i t_i - n_i r_i < e_i' t_i'-n_i'r_i') \land u_i') \lor
((e_i t_i - n_i r_i > e_i' t_i'-n_i'r_i') \land u_i) \lor$,
 $l_i''=l_i \lor l_i'$.
\end{definition}

\begin{definition}[Tile tensor summation]
Let $T=(E,S)$ be a tile tensor such that for a given index $i$ it holds that $u_i=false,r_i=1$. Then $T'=sum(T,i)$ is a tile tensor $T'=(E',S')$ computed as follows.
Let $E_1=sum(E,i)$. $E'$ is computed from $E_1$ by summing over the dimension $i$ of every tile $L$ of $E_1$ using the rotate-and-sum algorithms \cite{TILETENSORS}.
$S'$ is identical to $S$ except $n_i'=1$, and if 
$\forall_{j<i} t_j=1$ and $t_i$ is a power of $2$, then $r_i'=t_i$, else $u_i'=(t_i>1)$.
\label{def_tt_sum}
\end{definition}

\begin{remark}
The output tile tensor shape of tile tensor summation is due to the behaviour of rotate-and-sum algorithms as explained in \cite{TILETENSORS}. In environments where summing inside a tile can be performed differently, the shape might be different. Specifically, Some \gls{HE} systems support native multi-dimensional structure to the ciphertexts, allowing rotating a tile along one of its dimensions directly. This allows having replicated output for any dimension.
\end{remark}

\begin{remark}
The constraint $u_i=false$ in Definition~\ref{def_tt_sum} can be removed with some straightforward modifications. These details are omitted.
\end{remark}

\begin{definition}[Tile tensor replication]
Let $T=(E,S)$ be a tile tensor and $i$ be an index such that $n_i=1, r_i=1$, and $\forall_j u_j=false$. Then $T'=rep(T,i)$ is a tile tensor $T'=(E',S')$ computed as follows.
$E'$ is computed from $E$ by applying replication along dimension $i$ for 
on every tile $L$ of $E$ using the rotate-and-sum algorithms of \cite{TILETENSORS}.
$S'$ is identical to $S$ except $r_i'=t_i$.
\end{definition}

\subsection{Tile Tensor Glossary}
\label{glossary}
Below is a short summary of tile tensor terminology. 
\begin{itemize}
    \item {\em Tile} A tensor of some fixed shape, stored flattened inside a vector and operated on in \gls{SIMD} fashion.
    \item {\em Tile tensor } A data structure containing an {\em external tensor} as data and a {\em tile tensor shape} as meta data.
    \item {\em External tensor} A tensor with tiles as elements.
    \item {\em Tile shape} The shape of tiles in the external tensor.
    \item {\em Tile tensor shape } Meta data specifying the original shape, tile shape, and additional packing details.
    \item {\em Packed tensor } The tensor that will be the result of unpacking a tile tensor.
    \item {\em Original shape} The shape of the packed tensor.
\end{itemize}

\section{Optimizer algorithms}\label{sec:optalg}

\begin{algorithm}[ht!]
\caption{The optimizer operation}
\label{alg:optimizer}
\begin{algorithmic}[1]
    \Statex \textbf{Input:} $m$ - a JSON description of a machine learning model. 
    \Statex \qquad  $Req$ - a key-value dictionary of user requirements. 
    \Statex \qquad  $O$ - an ordered list of optimization targets e.g., `latency' and 'memory'. 
    \Statex \qquad  $Constraints$ - a a key-value dictionary of constraints. 
    \Statex \textbf{Output:} $p$ - an optimal packing configuration for $m$.
    \Procedure{Optimizer}{$m$, $Req$, $O$, $Cons$}
        \State $p = null$
        \State $pRes = null$
        \For{$conf \in \operatorname{GenerateConfs}()$}
            \State $packedM = \operatorname{Pack}(m, conf)$
            \State $res = \operatorname{Simulate/Eval}(packedM, m)$
            \For{$r \in Req.keys()$}
                \If{$res[r] > Req[r]$ }
                \State Continue with next $conf$
                \EndIf
            \EndFor
            \For{$r \in Constraints.keys()$}
                \If{$res[r] > Constraints[r]$ }
                \State Continue with next conf
                \EndIf
            \EndFor
            \For{$o \in O$}
                \If{$p != null$ \textbf{and} $res[o] >= pRes[o]$}
                \State Continue with next conf
                \EndIf
            \EndFor
            \State $(p, pRes) = (conf, res)$
        \EndFor
        \State \Return $p$
    \EndProcedure
\end{algorithmic}
\end{algorithm}

\begin{algorithm}[ht!]
\caption{GenerateConf}
\label{alg:genconf}
\begin{algorithmic}[1]
    \Statex \textbf{Input:} $m$ - a JSON description of a machine learning model. 
    \Statex \textbf{Output:} $l$ - A list of possible configurations. 
    \Procedure{GenerateConf}{$m$}
        \State $l = null$
        \State $n = \operatorname{GetMinimalMulDepth}(m)$
        \State $s = \operatorname{GetTileSize}(n)$
        \State $d = \operatorname{GetNumberOfTileTensorDims}(m)$
        \Statex
        \Comment e.g., depends on the existence of a convolutional layer in $m$.
        \State $c = \operatorname{GetSupportedConvModes}(m)$
        \State $ts = \operatorname{GetTileTensorShapes}(d, s)$ where
        \Statex
        \Comment the product of the shape denominators equals $s$
        \Statex
        \Comment the denominators are powers of two
        \State \Return $l$, a list of all element combinations from $ts$ and $c$
    \EndProcedure
\end{algorithmic}
\end{algorithm}

\section{Optimizer Accuracy}\label{sec:optres}

\begin{table}[ht!]
\begin{center}
\caption{Simulated time estimations for the configurations $(Conf_i)_{i=1..4}$ formatted as [tile shape - convolution mode (Section~\ref{subsec:conv_sequence})]: $[16,8,8,16,1]$-CWHFB, $[8,8,8,32,1]$-CWHFB, 
$[16,8,8,16,1]$-FWHCB, 
$[32,8,8,8,1]$-FWHCB, respectively. The acronyms CWHFB and FWHCB indicate the order of dimensions in the tile tensor. The deviation of the estimated times from the real times are reported in brackets.}
\label{tab_optimizer_1}
\begin{tabular}{clll}
   \textbf{Config} & \textbf{Inference}  & \textbf{Model enc.} & \textbf{Input enc.} \\ 
   & \textbf{(sec)} & \textbf{(sec)} & \textbf{(sec)} \\
\hline
$Conf_1$ & 4{,}232 (-11\%)&   1{,}509 (-11.5\%)& 162 (-6.8\%)\\
$Conf_2$ & 4{,}758 (-13.9\%)& 1{,}493 (-12.1\%)& 164 (-7.9\%)\\
$Conf_3$ & 4{,}927 (-18.1\%)& 1{,}680 (-11.5\%)& 177 (-6.8\%)\\
$Conf_4$ & 4{,}798 (-20\%)&   1{,}668 (-12.3\%)& 178 (-7.3\%)\\
\end{tabular}
\end{center}
\end{table}

The optimizer's simulator (Section \ref{section:optimizer}) estimates the time and memory usage for a given configuration option on a single CPU thread. For that, it relies on pre-benchmarked measures of the different HE operations. To assess the accuracy of these estimations, we performed the following experiment on HE-friendly AlexNet using encrypted model. We chose the four configuration options that achieved the lowest estimated latency when using local search (Section \ref{section:optimizer}) and compared the inference time and the encryption time of the input and the model between the simulation output and an actual run over encrypted data. 

Table \ref{tab_optimizer_1} summarizes the results. Our empirical tests show that the simulator provides relatively accurate time estimations for all four configurations. The average estimated time deviation is -15.8\%, -11.9\%, and -7.2\% for inference, model encryption, and batch input encryption, respectively. We note that the simulated storage matches the measured storage for all configurations, thus we do not include this data in Table \ref{tab_optimizer_1}.

\section{A proof of Lemma \ref{lem:conv1}}
\label{sec:proof}

\convlemma*

\begin{proof}
{\bf Multiplications.}
To compute the convolution, we need to multiply each of the $w_Ih_I$ elements of the input tensor with each of the $w_Fh_F$ elements of the filter, excluding edge cases that do not change the asymptotic behavior. Since each multiplication multiplies $s$ slots, we need only $\BigO(\ceil {w_I h_I w_F h_F / s})$ multiplications.

{\bf Rotations.}
Recall the output is of size $(w_I - w_F+1)(h_I - h_F+1)$. 
We map the $k$-th slot of different ciphertexts to elements of $I$ with indexes $k\ceil{\frac{w_I}{t_1}} \le x_o < (k+1)\ceil{\frac{w_I}{t_1}}$ and $k \ceil{\frac{h_I}{t_2}} \le y_o < (k+1)\ceil{\frac{h_I}{t_2}}$.
It is therefore enough to analyze the cost of computing the  convolution for $0 \le x_o < \ceil{\frac{w_O}{t_1}}$ and $0 \le y_o < \ceil{\frac{h_O}{t_2}}$, since computing the other elements of the output has no cost due to the SIMD feature.
It follows that a rotation is needed when $x_o + i \ge \ceil{\frac{w_I}{t_1}}$ or $y_o + j \ge \ceil{\frac{h_I}{t_2}}.$
This totals to $\BigO(w_F \ceil{\frac{w_I}{t_1}} + h_F \ceil{\frac{h_I}{t_2}} + w_Fh_F).$

{\bf Storage.}
Since we use $\BigO(s)$ slots of each ciphertext, the input can be encoded in $\BigO(w_Ih_I/s)$ ciphertexts.
\end{proof}

\section{Convolution operation counts}
\label{app:conv_perf}

It is easy to deduce the number of tiles from a tile tensor shape. The image tile tensor
$T_I[\frac{w_I\sim}{t_1},\frac{h_I\sim}{t_2},\frac{c}{t_3},\frac{b}{t_4},\frac{*}{t_5}]$ has
\[
\ceil{\frac{w_I}{t_1}}\ceil{\frac{h_I}{t_2}}\ceil{\frac{c}{t_3}}\ceil{\frac{b}{t_4}}
\]
tiles, and the filters tile tensor 
$T_F^l[w_F,h_F,\frac{*}{t_1},\frac{*}{t_2},\frac{c}{t_3},\frac{*}{t_4},\frac{f}{t_5}]$ has  
\[
w_F h_F \ceil{\frac{c}{t_3}} \ceil{\frac{f}{t_5}}
\] tiles.
We multiply every image tile with every filter tile that covers the same range of channels. This results with 
\[
\ceil{\frac{w_I}{t_1}}\ceil{\frac{h_I}{t_2}}\ceil{\frac{c}{t_3}}\ceil{\frac{b}{t_4}} w_F h_F \ceil{\frac{f}{t_5}}
\]
multiplications. 

In contrast, it is more difficult to compute the number of rotations. We can compute it exactly for the general case, using the following formulas.
The number of image tiles that we rotate due to the filter exceeding to the right, Equation ~\ref{rots_ex_right};  due to exceeding from below, Equation ~\ref{rots_ex_down}; and when both are exceeded, Equation ~\ref{rots_ex_both}. Additional rotations are required for summing over the channels dimension using rotate and sum, Equation~\ref{rots_sum}, and every second convolution layer the same amount again is needed for duplicating along this dimension.
 
 \begin{equation}
 \label{rots_ex_right}
     (w_F-1) \ceil{\frac{h_I}{t_2}}\ceil{\frac{c}{t_3}}\ceil{\frac{b}{t_4}}
 \end{equation}
 \begin{equation}
 \label{rots_ex_down}
 (h_F-1) \ceil{\frac{w_I}{t_1}}\ceil{\frac{c}{t_3}}\ceil{\frac{b}{t_4}}
 \end{equation}
 \begin{equation}
 \label{rots_ex_both}
 (w_F-1)(h_F-1)\ceil{\frac{c}{t_3}}\ceil{\frac{b}{t_4}}
 \end{equation}
 \begin{equation}
 \label{rots_sum}
 \ceil{\frac{w_I}{t_1}}\ceil{\frac{h_I}{t_2}}log t_3\ceil{\frac{b}{t_4}}\ceil{\frac{f}{t_5}}
 \end{equation}
 
 The above formulas are accurate for the general case, however, they over-count for certain special cases. These include the case where the filter is very large, and hence its valid positions do not cause it to exceed with its full length to neighboring tiles. Another case where the tile sizes $t_1$ or $t_2$ are $1$, and rotating them is redundant.

\end{document}